\documentclass[prd,nofootinbib,preprintnumbers]{revtex4}

\usepackage{latexsym}
\usepackage{amssymb}
\usepackage{amsmath}
\usepackage{xspace}
\usepackage{subfigure}

\usepackage{graphicx}

\usepackage [english] {babel}
\usepackage[dvips]{epsfig}
\numberwithin{equation}{section}

\DeclareMathOperator{\ctg}{ctg} \DeclareMathOperator{\tg}{tg}
\DeclareMathOperator{\ch}{ch} \DeclareMathOperator{\sh}{sh}
\DeclareMathOperator{\Sp}{Sp}

\newcommand{\comment}[1]{}

\newcommand{\hsp}{\hspace{0.15em}}

\newcommand{\thcr}{\vartheta_{\rm cr}}

\newcommand{\ds}{\displaystyle}

\newcommand{\cp}{{\cal{P}}}

\newcommand{\vk}{\varkappa}

\newcommand{\be}{\begin{equation}}
\newcommand{\ee}{\end{equation}}
\newcommand{\ba}{\begin{align}}
\newcommand{\ea}{\end{align}}
\newcommand{\bea}{\begin{eqnarray}}
\newcommand{\eea}{\end{eqnarray}}
\newcommand{\bw}{\begin{widetext}}
\newcommand{\ew}{\end{widetext}}

\newcommand{\ep}{{\varepsilon}}

\newcommand{\e}{{\rm e}}

\newcommand{\nn}{\nonumber}
\newcommand{\zt}{\dot{z}}
\newcommand{\ztt}{\ddot{z}}

\newcommand{\cd}{,\! \,}
\newcommand{\un}{\, ^1 \!}
\newcommand{\nul}{\, ^0\!}

\newcommand{\pp}{\, ... \,}

\newcommand{\eff}{\mathrm{eff}}

\newcommand{\Hh}{h}

\newcommand{\dq}{d^{{\kern 1pt}4} {\kern -1pt}q}

\newcommand{\kn}{{\kern 1pt}}
\newcommand{\akn}{{\kern -1pt}}

\newcommand{\vph}{\vphantom{{c'}^M}}
\newcommand{\vp}{\vphantom{{c}^M}}

\begin{document}

\title{Gravitational radiation in massless-particle collisions}
\author{Pavel
Spirin$^{1,2}$ and Theodore\,N.\,Tomaras$^{1}$
\thanks{\tt
salotop@list.ru, tomaras@physics.uoc.gr}}
\affiliation{\parbox{12cm}{\noindent{}$^{1}$\rule{0cm}{0.4cm}
Institute of Theoretical and Computational Physics, Department of
Physics,\\
\hphantom{m}  University of Crete, 70013 Heraklion, Greece;
\\ {}$^{2}${}  Department
of Theoretical Physics, Faculty of Physics, Moscow State
University,\\ \hphantom{m} 119899 Moscow, Russian Federation.}}

\date{\today}

\begin{abstract}
The angular and frequency characteristics of the gravitational
radiation emitted in collisions of massless particles is studied
perturbatively in the context of classical General Relativity for
small values of the ratio $\alpha\equiv 2 r_S/b$ of the
Schwarzschild radius over the impact parameter. The particles are
described with their trajectories, while the contribution of the
leading nonlinear terms of the gravitational action is also taken
into account. The old quantum results are reproduced in the zero
frequency limit \mbox{$\omega\ll 1/b$}. The radiation efficiency
$\epsilon \equiv E_{\rm rad}/2E$ {\it outside} a narrow cone of
angle $\alpha$ in the forward and backward directions with respect
to the initial particle trajectories is given by \mbox{$\epsilon
\sim \alpha^2$} and is dominated by radiation with characteristic
frequency \mbox{$\omega \sim {\mathcal O}\kn(1/r_S)$}.
\end{abstract}
\maketitle

\tableofcontents
\section{Introduction}
\label{intr}

The problem of gravitational radiation in particle collisions has
a long history and has been studied in a variety of approaches and
approximations. The interested reader may find a long and rather
comprehensive list of relevant references in \cite{GST}, where the
emitted gravitational energy, as well as its angular and frequency
distributions in ultra-relativistic {\it massive-}particle
collisions were computed. The condition imposed in \cite{GST} that
the radiation field should be much smaller than the zeroth order
flat metric restricted the region of validity of our approach to
impact parameters $b$ much greater than the inverse mass of the
colliding particles, and made our conclusions not applicable to
the massless case.

However, the problem of gravitational radiation in {\it
massless-}particle collisions is worth studying in its own right
and has attracted the interest of many authors in the past as well
as very recently. Apart from its obvious relevance in the context
of TeV-scale gravity models with large extra dimensions
\cite{GKST}, it is very important in relation to the structure of
string theory and the issue of black-hole formation in
ultra-planckian collisions \cite{ACV}, \cite{Dvali}. Nevertheless,
to the best of our knowledge, complete understanding of all facets
of the problem is still lacking. The emission of radiation in the
form of {\it soft} gravitons was computed in \cite{Weinberg} in
the context of quantum field theory, but in that computation the
contribution of the non-linear graviton self-couplings i.e. the
stress part of the energy-momentum tensor, was argued to be
negligible. The result of the quantum computation for
low-frequency graviton emission was reproduced by a purely
classical computation in \cite{Wbook}, due entirely to the
colliding particles and leaving out the contribution of the stress
part of the energy-momentum tensor.

In the pioneering work \cite{DEath} or its recent generalization
to arbitrary dimensions \cite{Herdeiro}, the special
case of collisions with vanishing impact parameter was studied, with emphasis on the
contribution to the radiation of the stress part of the energy momentum tensor, leaving out
the part related to the colliding particles themselves. In a more
recent attempt \cite{Taliotis} the metric was computed to second
order, but no computation of the radiation characteristics was
presented, apart from an estimate of the emitted energy based
essentially on dimensional analysis. More recently, a new approach
was put forward for the computation of the characteristics of the
emitted radiation \cite{GV}, based on the Fraunhofer approximation
of radiation theory. However, this method cannot be trusted at
very low frequencies $\omega \ll 1/b$ and, furthermore, it ignores
the non-linear terms of the gravitational action, which are
expected to be important in the high frequency regime. Thus, we believe it is
fair to conclude, that the issue of the frequency and angular
characteristics as well as the efficiency of gravitational
radiation in ultra-relativistic particle collisions is not
completely settled yet.

The purpose of the present paper is to extend the method used in
\cite{GST} to the study of gravitational radiation in collisions
of massless particles with center-of-mass energy $2E$ and impact
parameter $b$. The formal limit $m\to 0$ (or equivalently
$\gamma_{\rm cm} \to \infty$ for the Lorentz factor) of the
massive case leads to nonsensical answers for the radiation
efficiency, i.e. the ratio $\epsilon\equiv E_{\rm rad}/2E \sim
(r_S/b)^3 \gamma_{\rm cm}$ of the radiated to the available
energy, the characteristic radiation frequency $\omega\sim
\gamma^2_{\rm cm} /b$, or the characteristic emission angle
$\vartheta\sim 1/\gamma_{\rm cm}$. The whole set-up of the
computation in the massive case is special to that case and,
consequently, does not allow to extract safe conclusions related
to massless-particle collisions. In particular, the massive case
computation was performed in the lab frame, the choice of
polarization tensors was special to the lab frame, while, being
interested in ultra-relativistic collisions, we organized the
computation of the energy-momentum source in a power series of the
Lorentz factor $\gamma$. Here, we shall deal directly with
massless collisions in the center-of-mass frame and correct the
above inadequacies of our previous results. We shall study
classically the gravitational radiation in the collision of
massless particles using the same perturbative approach as in
\cite{GST}. The scattered particles will be described by their
classical trajectories, eliminating potential ambiguities in the
separation of the radiation field from the field of the colliding
particles, inherent in other approaches. Furthermore, at the level
of our approximation we shall take into account the contribution
of the cubic terms of the gravitational action to the radiation
source, which will be shown to be essential for the consistency of
our approach. Finally, the efficiency $\epsilon$ outside a narrow
cone in the forward and backward directions will be obtained as a
function of the only available dimensionless quantity $\alpha
\equiv 2r_S/b = 8GE/bc^4$, formed out of the four parameters $G,
E, b, c$, relevant to the problem at hand.

The rest of this paper is organized as follows: In Section 2 we
describe the model, our notation, the equations of motion and the
perturbative scheme in our approach. This is followed by the
computation in Section 3 of the total radiation amplitude, i.e.
the sum of the local and stress part. Section 4 focuses on the
study of the angular and frequency characteristics of the emitted
radiation in the most important regimes of the
emission-angle$-$frequency plane. . Furthermore, in a separate
subsection we compare the results of this paper to previous work
and verify that they are compatible in their common regime of
validity. Our conclusions are summarized in Section 5, while in
three Appendices the interested reader may find the details of
several steps of the computations and the proofs of basic
formulae, used in the main text.

\section{Notation -- Equations of motion}\label{eom}

The action describing the two massless particles and their gravitational interaction reads
\begin{equation} \label{action}
S=- \frac{1}{2} \sum \int  e (\sigma )\, g_{
\mu\nu}\!\left(z(\sigma)\vp\right)\dot{z}^{\mu}(\sigma)\,
\dot{z}^{\nu}(\sigma) \,d\sigma  -\frac{1}{\vk^2}\int R
\sqrt{-g}\, d^{\kn 4}\akn x\,,
 \end{equation}
where $e(\sigma)$ is the einbein of the trajectory $z^\mu(\sigma)$
in terms of the corresponding affine parameter $\sigma$,
$\vk^2=16\pi G$ and the summation is over the two particles. We
will be using unprimed and primed symbols to denote quantities
related to the two particles.

Variation of the einbeine gives for each particle the constraint
\begin{equation} g_{\mu\nu}\!\left(z(\sigma)\vp\right)\,
\dot{z}^{\mu}(\sigma)\, \dot{z}^{\nu}(\sigma)=0 \,.
\end{equation}
Using the $\sigma-$reparametrization invariance $\sigma \to
\tilde\sigma=\tilde\sigma(\sigma), \; e(\sigma) \to \tilde
e(\tilde \sigma) = e(\sigma) \,d\sigma/d\tilde\sigma $ we can
choose $e(\sigma)=$ constant. Furthermore, we can use the
remaining freedom of $\sigma$ rescalings to set $e=e'$ on the
particle trajectories. Finally, we can shift the affine parameters
to set $\sigma=0=\sigma'$ at the positions of closest approach of
the two particles. Before the collision the particle positions are
at negative $\sigma$ and $\sigma'$. They ``collide" when they are
at $\sigma=0=\sigma'$.

For identical colliding particles in the center-of-mass frame we
can choose $\sigma '=\sigma $ and, consequently, $e=e'$. With the
gauge choice $e={\rm constant}$, the two einbeine are finally
determined by the condition
 \begin{equation}
 \label{eew}
\sqrt{s}=E+E'=\int T^{00}(x)\, d^{\kn 3}  \mathbf{x}\,,
 \end{equation}
from which
 \begin{equation} \label{eeq}
e=\sqrt{s}/2=E\,,
 \end{equation}
with $E$ the energy of each colliding particle.

Thus, the particles move on null geodesics, while variation of $z^{\mu}$ leads to the particle equation of motion:
\begin{align}
\label{peom}
 \frac{d}{d\sigma } \left(g_{\mu\nu}\zt^\nu\right)
 = \frac{1}{2}\,g_{\lambda\nu,\kn\mu} \zt^\lambda  \zt^\nu
\end{align}
and similarly for $z'^\mu$. At zeroth order in the gravitational
interaction, the space-time is flat and the particles move on
straight lines with constant velocities, i.e.\kn\footnote{The
upper left index on a symbol labels its order in our perturbation
scheme.}
$$  \nul g_{\mu\nu}=\eta_{\mu\nu} \,; \qquad \nul \dot{z}^\mu \equiv u^\mu=(1,0,
0,1)\,, \qquad\nul \dot{z}'^\mu \equiv u'^\mu=(1,0,0,-1)\,.$$ The
particle energy-momentum is defined by $T^{\mu\nu} \equiv
(-2/\sqrt{-g}) \,\delta S/\delta g_{\mu\nu}$\,, i.e. for each
particle
\begin{equation}
\label{Tmn}
T^{\mu\nu}(x)= e \int\frac{\zt^{\mu}
\zt^{\nu}\,\delta(x-z(\sigma ))}{ \sqrt{-g}} \,d\sigma \,.
\end{equation}
At zeroth order, in particular, it is given by
\begin{equation} ^0T^{\mu\nu}=e \int\frac{u^{\mu}
u^{\nu}\,\delta(x-z(\sigma ))}{ \sqrt{-g}} \,d\sigma
\end{equation}
and is the source of the first correction $h_{\mu\nu}$ of the
gravitational field. Given that $\nul \,T_{\mu\nu}$ is traceless,
the perturbation $h_{\mu\nu}$ satisfies for each particle
separately the equation
\begin{align} \label{grsola}
\partial^2 h_{ \mu\nu}=-\vk \nul\, T_{ \mu\nu}\,,
\end{align}
whose solution in Fourier space is
\begin{align}
\label{hh'}
 h_{ \mu\nu}(q)=
 \frac{2\pi \vk}{q^2+i {\kern 1pt}0  {\kern 1pt}q^0}\,\e^{iqz(0)}\delta(qp) \,p_{ \mu}p_{ \nu}\qquad \text{and}  \qquad h'_{ \mu\nu}(q)=
 \frac{2\pi \vk}{q^2+i {\kern 1pt}0 {\kern 1pt} q^0}\,\e^{iqz'(0)}\delta(qp')\,
 p'_{ \mu}p'_{ \nu}\, ,
 \end{align}
where $p^{ \mu}=e\kn u^{ \mu}$, $p'^{ \mu}=e\kn u'^{ \mu}$, while
$z^{\mu}(0)= ( 0, b/2, 0,0 )$ and $z'^\mu(0)= ( 0, -b/2, 0, 0 )$.
Since the particle momenta satisfy $p^2=0=p'^2$, the consistency
conditions $h^\mu_{\;\mu}=0=h'^\mu_{\;\;\mu}$ are also satisfied
to this order.

In coordinate representation they are
 \begin{align}
 \label{AShmn}
& h_{\mu\nu}(x) = -\frac{\vk  {\kern 1pt} e {\kern 1pt}
u_{\mu}u_{\nu}}{(2\pi)^{3} } \int \frac{dq^z d^{\kn 2} \mathbf{q}
}{\mathbf{q}^2} \,\e^{-i q^z (t-z)} \e^{i\mathbf{q}[{\bf r}-{\bf
b}/2]} =  -\vk {\kern 1pt}e{\kern 1pt} u_{\mu} u_{\nu} {\kern 1pt}
\delta(t-z)\, \Phi(|{\bf r}-{\bf b}/2|) \, \nn
\\ & h'_{\mu\nu}(x) = -\frac{\vk {\kern 1pt} e  {\kern 1pt} u'_{\mu}\,u'_{\nu}}{(2\pi)^{3}
} \int  \frac{dq^z d^{\kn 2} \mathbf{q} }{\mathbf{q}^2}\, \e^{i
q^z (t+z)} \e^{i\mathbf{q}[{\bf r}+{\bf b}/2]} =-\vk {\kern 1pt} e
{\kern 1pt} u'_{\mu} u'_{\nu}  {\kern 1pt}\delta(t+z) \,\Phi(|{\bf
r}+{\bf b}/2|)\,,
\end{align}
where $\Phi$ is the $2-$dimensional Fourier transform of
$1/q^2$:
 \begin{align}
 \label{iii0}
\Phi(r) \equiv \frac{1}{(2\pi)^{2} }  \int \frac{ d^{\kn 2}
\mathbf{q} }{\mathbf{q}^2}\: \e^{-i\mathbf{q} \mathbf{r}} =
    -\ds\frac{1}{2\pi} \ln \frac{r}{r_0}
\end{align}
with ${\bf r}=(x,y)$ and ${\bf b}=(b,0)$ is the position and impact vector, respectively, in the transverse $x-y-$plane and $r_0$ an arbitrary constant with dimensions of length.

Write for the metric $g_{\mu\nu}=\eta_{\mu\nu}+\vk (h_{\mu\nu}+h'_{\mu\nu})$ and substitute in (\ref{peom}) to obtain for the first correction of the trajectory of the unprimed particle the equation
\begin{align}
\label{peom1} \un \ztt_{\mu} ( \sigma )
=-\vk\left(h'_{\mu\nu,\lambda}-\frac{1}{2}\,h'_{\lambda\nu,\kn\mu}\right)\nul\zt^\lambda
\nul\zt^\nu.
\end{align}
The interaction with the self-field of the particle has been omitted and $h'_{\mu\nu}$ due to the primed particle is evaluated at the location of the unprimed particle on its unperturbed trajectory.

We substitute (\ref{hh'})  into (\ref{peom1}) to obtain
\begin{align}
\label{sc6c}
    \un \ztt^{\mu}( \sigma )  =
 \frac{2 i \kn e \vk^2}{(2 \pi)^{3}
 }
 \int \dq
\frac{\delta(qu')}{ q^2  }\, \e^{-iqb} \e^{-i(qu)  \sigma } \left[
(qu)\, u'^{\mu} -   q^{\mu} \vph\right].
\end{align}
Integrating it over $\sigma $, the first-order correction to velocity
is given by
\begin{align}
\label{sc6c_b1} \un \zt^{\mu}( \sigma  )  = -\frac{2 \kn e
\vk^2}{(2 \pi)^{3}
 }
 \int \dq
\frac{\delta(qu')}{ q^2}\,\e^{-iqb} \e^{-i(qu)  \sigma } \left[
u'^{\mu} -\frac{  q^{\mu}} { (qu)}\right]+C^{\mu}.
\end{align}
The integration constants  $C^{\mu}$ are chosen $C^0=0=C^z$ and $C^x=e \vk^2 \Phi'(b)/2$ in order to satisfy the initial conditions $^1\!\dot z^\mu(\sigma=-\infty)=0$.

Thus, the components of $\un \zt^{\mu}( \sigma)$ are
\begin{align}
\label{sc6c_pp0}
 &\un \zt^{0}( \sigma  )  =  \frac{  e \vk^2}{(2 \pi)^{3}
 }
 \int dq^0\,
\frac{d^{\kn 2} \mathbf{q}}{ \mathbf{q}^2}\,\e^{i\mathbf{qb}}
\e^{-2iq^0  \sigma } =\frac{1}{2}\, e
\vk^2 \Phi(b)\, \delta( \sigma ) \nn \\
 & \un \zt^{z}( \sigma  )=-\frac{1}{2}\, e
\vk^2 \Phi(b)\, \delta( \sigma ) \nn \\
&\un \zt^{x}( \sigma  )  = -\frac{  e \vk^2}{(2 \pi)^{3}
 }
 \int\frac{ dq^0}{  q^0}
\frac{d^{\kn 2} \mathbf{q}}{ \mathbf{q}^2}\,\e^{i\mathbf{qb}}
\e^{-2i q^0 \sigma }
     q^{x} + C^x= e \vk^2 \, \Phi'(b)\, \theta(\sigma) \nn \\
 &\un \zt^{y}( \sigma  )=0\,.
\end{align}

Making use of the formulae \cite{GS}
$$\frac{1}{[x+i0]^n} =\frac{1}{x^n} -i\pi \frac{(-1)^{n-1}}{(n-1)!}\:\delta^{(n-1)}(x)\,,
 \qquad\qquad   \mathcal{F}\left[\frac{1}{(x+i0)^n}\right](k)= 2\pi\frac{ (-i)^n}{(n-1)!} \,[k\,\theta(-k)]^{n-1}\,,$$
satisfied by the distributions $(x+i0)^{-n}$ and their Fourier
transform, respectively, we can express $\un \zt^{\mu}( \sigma)$
collectively in the following useful form
\begin{align}
\label{z1m}
\un \zt^{\mu}( \sigma  )  = -\frac{2 e \vk^2}{(2 \pi)^{3}
 }
 \int \dq
\frac{\delta(qu')}{ q^2}\,\e^{-iqb}\, \e^{-i(qu)  \sigma }  \left[
(qu){\kern 1pt} u'^{\mu} -  q^{\mu} \vph
\right]\frac{1}{(qu)+i0}\, ,
\end{align}
which vanish for all \mbox{$\sigma <0$}. Indeed, the massless
particle trajectories should remain undisturbed before the
collision.

Finally, we integrate (\ref{z1m}) and fix the integration
constants so that $\un z^\mu(\sigma)$ is regular and satisfies
\mbox{$\un z^\mu(\sigma<0)=0$}. We end up with
\begin{align}\label{sc6c_b4}
\un z^{\mu}( \sigma  )  = -\frac{2 i e \vk^2}{(2 \pi)^{3}}\int \dq
\frac{\delta(qu')}{q^2  }\,\e^{-iqb}\, \e^{-i(qu)  \sigma } \left[
(qu){\kern 1pt} u'^{\mu} -  q^{\mu} \vph
\right]\frac{1}{[(qu)+i0]^2} \, ,
\end{align}
or, equivalently, in components
\begin{align}
\label{ddd1}
& \un z^{0}( \sigma )= \frac{1}{2}\, e \vk^2 \Phi(b)\, \theta( \sigma )
=-\un z^{z}( \sigma )\nn \\
& \un z^{x}( \sigma )=e \vk^2 \Phi'(b)\, \sigma {\kern 1pt}\theta(
\sigma )\,.
\end{align}
From these it is straightforward to reproduce the leading order expressions of the two well-known facts about the geodesics in an Aichelburg-Sexl metric, namely
\begin{itemize}
    \item The {\it time delay} at the moment of shock equal $$ \Delta t =   e \vk^2 \Phi(b)=8{\kern 1pt} GE\,\ln \frac{b}{r_0} \,;$$
    \item  The {\it refraction} caused by the gravitational interaction by an
    angle
$$ \alpha =   e \vk^2  \, |\Phi'(b)|=\frac{8{\kern 1pt}GE}{b} $$
in the direction of the center of gravity.
\end{itemize}

Clearly, similar expressions to the above are obtained for the primed particle trajectory. For the perturbation $\un z'^\mu(\sigma)$, in particular, we have
\begin{align}
\label{z'1m} \un z'^{\mu}( \sigma)  = -\frac{2 i e \vk^2}{(2
\pi)^{3}}\int \dq  \frac{\delta(qu)}{q^2  }\,\e^{+iqb}\e^{-i(qu')
\sigma }  \left[ (qu'){\kern 1pt} u^{\mu} -  q^{\mu} \vph
\right]\frac{1}{[(qu')+i0]^2} \, .
\end{align}

{\it To summarize: We have obtained the first order corrections
$h_{\mu\nu}(x)$ of the gravitational field, sourced by the straight
zeroth-order trajectories of two colliding massless particles. It is
identical with the leading term of the Aichelburg-Sexl metric
describing the free particles and it can be shown to coincide with the limit $m\to 0$
of the corresponding field due to massive particles. The perturbations $\un z^{\mu}( \sigma)$ and $\un z'^{\mu}( \sigma)$ of the trajectories of the colliding particles in the center-of-mass frame and with impact parameter $b$ were also computed. Finally, the known expressions \cite{Dray87} for the time delay $\Delta t$ and the leading order in $r_S/b \ll 1$ scattering angle $\alpha$ were reproduced.}

As will be shown in the next section, the arbitrary scale $r_0$ in
the expressions for $h_{\mu\nu}$ and $h'_{\mu\nu}$ disappears, as
it ought to, from physical quantities such as the gravitational
wave amplitude or the frequency and angular distributions of the
emitted energy.

\section{Radiation amplitude}

We proceed with the computation of the energy-momentum source of
the gravitational radiation field. The gravitational wave source
has two parts. One is the particle energy-momentum contribution,
localized on the accelerated particle trajectories given in the
previous section. The other is due to the non-linear
self-interactions of the gravitational field spread over
space-time. One should keep in mind that we are eventually
interested in the computation of the emitted energy, given by
(\ref{fr_di}). It involves projection of the energy-momentum
source on the polarization tensors and imposing the mass shell
condition on the emitted radiation wave-vector. Thus, whenever
convenient, we shall simplify the expressions for the Fourier
transform of the energy-momentum source by imposing the on-shell
condition $k^{\kn2}=0$, as well as by projecting it on the two
polarizations.

\subsection{Local source}

We start with the direct particle contribution to the source of
radiation. We call it ``local", because, as mentioned above, it is
localized on the particle trajectories. The first order term in
the expansion of (\ref{Tmn}) is
 \begin{align}
\label{Tmn1}
\un T_{\mu\nu}(x)= e\int  d\sigma \, & \left[ 2\un \zt_{(\mu}
u_{\nu)} +2 \vk u^{\lambda}h'_{\lambda
(\mu} u{\vphantom{h'_\lambda}}_{\nu)} - u_\mu u_\nu (\!\un {z}\cdot
\partial) \right] \delta^4 \!\!{\kern 1pt}\left(x-\! \! \nul z(\sigma )\right)\, ,
\end{align}
where $z^\mu$ is evaluated at $\sigma $ and $ h'_{\mu\nu}$ is
evaluated at $\nul z^\mu(\sigma ) $. Its Fourier transform is
\begin{align}
\label{Tmn1ofk}
\un T_{\mu\nu}(k)=\e^{ikz(0)} e \int d \sigma  \, \e^{i(ku)
 \sigma }\left[2 u_{( \mu} \un \dot{z}_{ \nu)}+2
\vk u^{ \lambda}h'_{ \lambda (
\mu}u{\vphantom{h'_\lambda}}_{\nu)}+ i(k\cdot\!\!\un z)\,
u_{\mu}u_{\nu}\right].
\end{align}
Similarly for the primed particle with $u$ replaced by $u'$.

Introducing the momentum integrals
\begin{align}
I\equiv \frac{1}{(2\pi)^{2}} \int \frac{ \delta(qu')\,
\delta(ku-qu)\,\e^{-i\kn(qb)}}{q^2}\;\dq \,, \qquad I_{\mu}\equiv
\frac{1}{(2\pi)^{2}} \int \frac{ \delta(qu')\, \delta(ku-qu)\,
\e^{- i\kn(qb)}}{q^2} \;q_{\mu} \; \dq \,, \nn
\end{align}
the first-order correction to the source
becomes\kn\footnote{Terms, coming from the integration constants
in (\ref{z1m}) and (\ref{sc6c_b4}), contain $\delta(qu)$ and lead
to the extra terms proportional to $\delta(ku)$ and $\delta'(ku)$.
With the on-shell condition $k^2=(ku)(ku')-k_{\bot}^2=0$ the
latter is equivalent to $k^\mu=0$ and these terms do not
contribute in the subsequent $d^{3}k-$integration. }
\begin{align}
\label{Tmn1B}
 \un T_{\mu\nu}=2{\kern 1pt} e^2 \vk^2 \e^{ikz(0)}
 \frac{1}{ (ku)} \left[ u_{\mu} u_{\nu}
\left( ku'I-\frac{k I}{ k u}\right)+2{\kern 1pt}  u_{( \mu} I_{
\nu)} \right]
 , \qquad  \un T'_{\mu\nu}=\left.\un T_{\mu\nu}\vp\right|_{ u \leftrightarrow
 u', b^{\kn\mu} \to
-b^{\kn \mu} }\,.
\end{align}

{\it Note that the integrals $I$ and $I_\mu$ contain one massless Green's function. This is in accordance with the fact that $\un T_{\mu\nu}$, expressed through them, is the source of radiation from the colliding particles. }
$I$ and $I_\mu$ are computed in Appendix \ref{Local_ints}. They are
\begin{align}
I=-\frac{1}{2} \,\Phi(b)   \,, \qquad I_{\mu}=-
\frac{(ku)\,\Phi(b)}{4}\,u'_\mu+i\frac{\Phi'(b)}{2b}\, b_\mu
\end{align}
and upon substitution into (\ref{Tmn1B}) lead to
\begin{align}
\label{gg1D}
 \un T_{\mu\nu}=-2 {\kern 1pt}  e^2 \vk^2 \e^{i(kb)/2}
  \biggl[ \Phi(b) \,u'_{(\mu} u^{\vphantom{\prime}}_{\nu )} +
  \frac{ (ku')\,\Phi(b)}{2 (ku)}\,u_{\mu}u_{\nu} +i\frac{\Phi'(b)  \, \sigma^{(u)}_{\mu\nu}}{b  \,(k
u)^2 } \biggr]
\end{align}
with $ \sigma^{(u)}_{\mu\nu} \equiv (kb)\, u_{\mu}u_{\nu}
-2(ku)\,u_{( \mu}b_{ \nu)}$. Similarly
\begin{align}
\label{gg1D'}
 \un T'_{\mu\nu}=-2{\kern 1pt}  e^2 \vk^2 \e^{-i(kb)/2}
\biggl[  \Phi(b) \,u'_{( \mu} u^{\vphantom{\prime}}_{ \nu)}+
  \frac{ (ku)\,\Phi(b)}{2 (ku')}\,u'_{\mu}u'_{\nu} -i\frac{\Phi'(b)  \, \sigma^{(u')}_{\mu\nu}}{b \, (k
u')^2 } \biggr]
\end{align}
for the contribution of the primed particle, obtained from $\un
T_{\mu\nu}$ by the substitution $b^{\kn\mu} \to -b^{\kn\mu}, u^\mu
\leftrightarrow u'^{\mu}$.

Eventually, $ \un T_{\mu\nu}$ and $ \un T'_{\mu\nu}$ will be contracted with the polarization vectors $e_1$ and $e_2$, we will construct in the next section. They have zero time component and, therefore, satisfy $e_1 \cdot u'=- e_1 \cdot u$ and $e_2 \cdot
u'=- e_2 \cdot u$. Thus, one may {\it effectively} replace in the energy momentum tensor $u'_\mu$ by $ -u_\mu$ when they are not contracted, to obtain
\begin{align}
\label{T1eff}
 \un T_{\mu\nu}=-2 {\kern 1pt} e^2 \vk^2 \e^{i(kb)/2}
 \biggl[ -  \Phi(b) \,u_{\mu} u_{\nu} +
  \frac{ (ku')\,\Phi(b)}{2 {\kern 1pt} (ku)}\,u_{\mu}u_{\nu} +i\frac{\Phi'(b)  \, \sigma^{(u)}_{\mu\nu}}{b\,  (k
u)^2 } \biggr]
\end{align}
and
\begin{align}
\label{T'1eff}
 \un T_{\mu\nu}^{\prime} =-2{\kern 1pt}  e^2 \vk^2 \e^{-i(kb)/2}
\biggl[ -  \Phi(b) \,u_{\mu} u_{\nu}+
  \frac{ (ku)\,\Phi(b)}{2{\kern 1pt}  (ku')}\,u_{\mu}u_{\nu}-i\frac{\Phi'(b)  \, \bar{\sigma}^{(u)}_{\mu\nu}}{b\,  (k
u')^2 }\biggr]\,,
\end{align}
where $\bar{\sigma}^{(u)}_{\mu\nu}  \equiv (kb)\,
u_{\mu}u_{\nu} +2{\kern 1pt} (ku')\,u_{( \mu}b_{ \nu)} $.

\subsection{Non-local stress source}

The contribution to the source at second-order coming from
the expansion of the Einstein tensor reads \cite{GST}
\begin{align}
S_{\mu\nu} (\Hh) =&\,
{\Hh}_\mu^{\lambda \cd \rho }(\Hh_{\nu \rho  \cd \lambda } -
\Hh_{\nu \lambda \cd \rho }) +\Hh^{\lambda \rho }(\Hh_{\mu \lambda
\cd \nu \rho }+ \Hh_{\nu \lambda \cd \mu \rho }- \Hh_{\lambda \rho
\cd \mu\nu}- \Hh_{\mu\nu
\cd \lambda \rho }) -  \nn\\
  -&\frac{1}{2}\, \Hh^{\lambda \rho }_{\quad \cd \mu } \Hh_{\lambda \rho  \cd
\nu }-\frac{1}{2}\,\Hh_{\mu\nu} \partial^2 \Hh +
\frac{1}{2}\,\eta_{\mu\nu}\left(2\Hh^{\lambda \rho }\partial^2
\Hh_{\lambda \rho }-\Hh_{\lambda \rho  \cd \sigma} \Hh^{\lambda
\sigma \cd \rho }+\frac{3}{2}\, \Hh_{\lambda \rho  \cd \sigma}
\Hh^{\lambda \rho \cd \sigma}\right).\nn
\end{align}
It contains products of two first-order fields. Thus, it is not localized, hence its name ``non-local". It is also called ``stress", being part of the stress tensor of the gravitational field.

Upon substitution of $h_{\mu\nu}$ and $\un z^\mu(\sigma)$ of the previous section in the above expression we obtain for the Fourier transform of $S_{\mu\nu}$
\begin{align}
S_{\mu\nu}(k) = \vk^2 e^2 \e^{i (kb)/2} & \left[ (ku')^2 u_\mu
u_\nu  J+ (ku)^2 u'_\mu  u'_\nu  J+4{\kern 1pt} J_{\mu\nu}
+4{\kern 1pt} (ku')\,u_{(\mu } J_{\nu )}-4{\kern 1pt}
(ku)\,u'_{(\mu } J^{\vphantom{\prime}}_{\nu )} + \right. \nn
\\ &\left.+2 {\kern 1pt} u_{(\mu } u'_{\nu )}\left(2{\kern 1pt} (kJ) -(ku)(ku')\,J-2\Sp J
\vph\right) \right]
\nonumber
\end{align}
in terms of the integrals
\begin{align}
\nonumber J_{\mu _1\,\pp \mu _l}(k)\equiv \frac{1}{(2\pi)^{2}}
\int \frac{ \delta(qu')\, \delta(ku-qu)\, \e^{- i(qb)}}{q^2
(k-q)^2} \;q_{\mu _1}\pp q_{\mu _l} \; \dq
\end{align}
($l=0, 1, 2$). We use the definition $\Sp J\equiv \eta^{\kn
\mu\nu} J_{\mu\nu}$, while we have omitted the terms proportional
to $\eta_{\mu\nu}$ as well as the longitudinal ones proportional
to $k_\mu$ or $k_\nu$ in anticipation of the fact that they will
eventually vanish, when contracted with the radiation polarization
tensors. Finally, as in the case of $\un T_{\mu\nu}$ one can {\it
effectively} substitute $u'_\mu  \to -u_\mu $ to obtain:
\begin{align}
\label{Smn}
S_{\mu\nu} (k) = \vk^2 e^2 \e^{i
(kb)/2}   \left[\!\left( 4 \Sp J-4{\kern 1pt} (kJ) +\left[\vp(ku')+(ku)\right]^2
J  \right)u_\mu u_\nu  +4J_{\mu\nu}
+4\left[\vphantom{d^d_g}(ku')+(ku)\right]u_{(\mu } J_{\nu )}\vph
\right].
\end{align}

{\it Note} that $J_{\mu_1\ldots}$ contain the product of two graviton Green's functions, which signals the fact that $S_{\mu\nu}$ is due to radiation from ``internal graviton lines" in a Feynman graph language, through the cubic graviton interaction terms. It will be explicitly demonstrated below that in the zero frequency limit the contribution of $S_{\mu\nu}$ in the emitted radiation is negligible, as argued in \cite{Weinberg}. Nevertheless, it will become clear that it contributes significantly at high frequencies and, as will be shown next, it plays an important role in the cancellation of the $r_0$ dependence in physical quantities.

\subsection{Cancellation of the arbitrary scale $r_0$}

As anticipated, in this subsection we will demonstrate explicitly
that the arbitrary scale $r_0$ disappears from the final
expression of the total contribution to the source $\un
T_{\mu\nu}+\un T\lefteqn{\smash{^{\prime}}}_{\!\mu\nu}+S_{\mu\nu}$
of the gravitational radiation. As will become clear below, the
local and stress parts of the source each depends on $r_0$, but
their sum is $r_0-$independent and finite. According to their
expressions in (\ref{T1eff}) and (\ref{T'1eff}), $\un T_{\mu\nu}$
and $\un T\lefteqn{\smash{^{\prime}}}_{\!\mu\nu}$ depend on $r_0$
through $\Phi(b)$, while $S_{\mu\nu}$ depends on $r_0$ through
terms proportional to $\hat{K}_{-1}(\zeta)$ (with no extra factors
$\zeta$) in the expressions of $J, J_\mu$ and $J_{\mu\nu}$,
evaluated in Appendix \ref{Local_ints}. All these unphysical terms
will be shown to cancel out and will end up with expressions
(\ref{tau+}) and (\ref{taux}) for the total energy-momentum source
for the two polarizations separately\kn\footnote{The reader, who
is not interested in the details, may go directly to these
formulae for the total source.}.

We proceed in steps:

\textbf{1.} Split $S_{\mu\nu} = S^I_{\mu\nu} + S^{I\akn
I}_{\mu\nu}$ with\kn\footnote{The integrals $k\cdot J$ and $\Sp J$
are singled-out, because they can be computed exactly. See
Appendix \ref{Local_ints}.}
\begin{align}
\label{nsplit}
& S^I_{\mu\nu} \equiv 4
\vk^2 e^2 \e^{i (kb)/2}  \!\left(\Sp J
 -(kJ)\vph \right) u_\mu  u_\nu \nn \\
&  S^{I\akn I}_{\mu\nu} \equiv \vk^2 e^2 \e^{i (kb)/2} \left[
\left[\vp(ku')+(ku)\right]^2 \,J \,u_\mu  u_\nu  \! +4J_{\mu\nu}
+4\left[\vphantom{d^d_g}(ku')+(ku)\right]u_{(\mu } J_{\nu )}\vph
\right].
\end{align}
Using \eqref{kJ} and \eqref{SpJ}, $ S^I_{\mu\nu}$ becomes
\begin{align}
\label{nsplit1}  S^I_{\mu\nu} =-
\vk^2 e^2 \Phi(b)\,\e^{i (kb)/2}  \!\left(  \e^{-i (k b)}+
 1\vp \right) u_\mu  u_\nu =-2{\kern 1pt} \vk^2 e^2 \Phi(b)\cos \frac{k\cdot b}{2}\, u_\mu  u_\nu\,.
\end{align}

\textbf{2.} Similarly, it is convenient to split the local source $\un T_{\mu\nu}+ \un T^{\prime}_{\mu\nu}$
(\ref{T1eff}, \ref{T'1eff}) as:
\begin{align}
\label{TI-II-III}
& T^{I}_{\mu\nu}=  e^2 \vk^2  \left[\e^{i(kb)/2} +\e^{-i(kb)/2}\right] \Phi(b) \,u_{\mu}u_{\nu}
=2{\kern 1pt}  \vk^2 e^2 \Phi(b)\cos \frac{k\cdot b}{2}\, u_\mu  u_\nu\, \nn \\
& T^{I\akn I}_{\mu\nu}=- e^2 \vk^2\frac{ \Phi(b)}{2}
\left[\e^{i(kb)/2}
\frac{ (ku') }{ (ku)} +\e^{-i(kb)/2} \frac{ (ku)}{ (ku')}  \right] u_{\mu}u_{\nu} \nn \\
& T^{I\!I\!I}_{\mu\nu}=- i e^2 \vk^2   \frac{\Phi'(b)}{b}
\biggl[\e^{i(kb)/2}
 \frac{ \sigma^{(u)}_{\mu\nu}}{ (ku)^2 } - \e^{-i(kb)/2} \frac{ \bar{\sigma}^{(u)}_{\mu\nu}}{ (ku')^2 }
 \biggr].
\end{align}

Thus, $T^I_{\mu\nu} + S^I_{\mu\nu}=0$.

\textbf{3.} The remaining stress contribution $ S^{I\akn
I}_{\mu\nu}$ is a linear combination of $J$, $J_\mu $ and
$J_{\mu\nu}$\kn, which have been computed in Appendix
\ref{Local_ints}. Taking, as above, into account the fact that
they will eventually be contracted with the polarization vectors
and that we shall set $k^2=0$ in the integral for the radiation
energy and momentum we are interested in, they
are\kn\footnote{Note that we use non-standard symbols for the
modified Bessel functions, namely $\hat{K}_{\nu}(z) \equiv
K_{\nu}(z)\,z^{\nu}$. In this notation the differentiation rule
reads $\hat{K}'_{\nu}(z)=-z \hat{K}_{\nu-1}$ for any $\nu$, while
the zero-argument limit is $\hat{K}_{\nu}(0)=2^{\nu-1}
\Gamma(\nu)$ for $ \nu>0$. }:
\begin{align}
&J = \frac{ b^{\kn 2}}{8\pi}\int\limits_0^1 dx \,\e^{-i (k b) x}
\hat{K}_{-1} \!\left(k_\perp b \sqrt{x(1-x)}
\right),  \nn \\
& J_\mu ^{\eff} = \frac{ b^{\kn 2}}{8\pi} \int\limits_0^1 dx
\;\e^{-i (k  b) x} \left[N^{\eff}_\mu    \hat{K}_{-1}
\!\left(\zeta \right)+ i \frac{b_\mu  }{b^{\kn 2}}\,{K}_{0}\!
\left(\zeta  \right) \right], \qquad N_\mu ^{\eff} \equiv
-\frac{1}{2}\left[\vph x(ku')+(1-x) (ku) \right]
u_\mu  \nn \\
&  J_{\mu\nu}^{\eff} = \frac{ 1}{8\pi} \int\limits_0^1
dx\;\e^{-i(kb)x} \left[ b^{\kn 2} N^{\eff}_{\mu} N^{\eff}_{\nu}
\hat{K}_{-1}(\zeta)+
 \left( 2iN^{\eff}_{(\mu } b_{\nu )}  - u_{\mu}u_{\nu} \vph \right)\! {K}_{0}( \zeta )-
\frac{b_{\mu}b_{\nu}}{b^{\kn 2}}\hat{K}_{1}( \zeta ) \right]\,.
\nn
\end{align}
Having anticipated that the dangerous terms for divergence and
$r_0-$dependence are the ones which contain the integral of $\hat
K_{-1}(\zeta)$ with $\zeta \equiv k_\perp b\sqrt{x(1-x)}$, since
according to Appendix \ref{Local_ints} lead to
$\Phi(b)$\kn\footnote{The integral containing the hatted Macdonald
of index ${-1}$, which near $x=0, 1$ behaves as $\hat{K}_{-1}
\!\left(k_\perp b \sqrt{x(1-x)} \right)\sim [x(1-x)]^{-1}$,
diverges logarithmically at both ends of the integration region.
In Appendix \ref{Local_ints} it is shown that this logarithmic
behavior is related to the one of $\Phi$ (Eqns. (\ref{kJ},
\ref{SpJ})). Alternatively, one could regularize these divergent
integrals by shifting the index of all Macdonald functions by
$0<\epsilon \ll 1$, which makes all $x-$integrations convergent,
and take the limit $\epsilon \to 0$ in the very end of the
computation. } it is natural to treat separately the terms in
$S^{I\akn I}_{\mu\nu}$ which contain $\hat K_{-1}$, from the ones
which contain $K_0$ or $\hat K_1$. Thus, in a suggestive notation,
we split: $S^{I\akn I}_{\mu\nu}=
S_{\mu\nu}^{(-1)}+S_{\mu\nu}^{{\kern 1pt}(0,1)}$ with
\begin{align}
\label{nsplit3} & S_{\mu\nu}^{(-1)}   \equiv 2{\kn} G b^{\kn 2}
e^2 \e^{i (kb)/2} \int\limits_0^1 dx \,\e^{-i (k b) x}
\hat{K}_{-1} \!\left( \zeta \right) \left[
\left[\vp(ku')+(ku)\right]^2  u_\mu u_\nu +4N^{\eff}_{\mu}
N^{\eff}_{\nu} +4\left[\vphantom{d^d_g}(ku')+(ku)\right]u_{(\mu }
N^{\eff}_{\nu )}\vph \right]
 \nn \\  &  S_{\mu\nu}^{{\kern 1pt}(0,1)}   \equiv  8{\kern 1pt} G e^2 \, \e^{i
(kb)/2} \!\! \int\limits_0^1 \! dx \,\e^{-i (k b) x} \!\left[
\left( 2iN^{\eff}_{(\mu } b_{\nu )}  - u_{\mu}u_{\nu} +i
\left[\vphantom{d^d_g}(ku')+(ku)\right] u_{(\mu } b_{\nu )}\vph
\right)\!
 {K}_{0}( \zeta )- \frac{b_{\mu}b_{\nu}}{b^{\kn 2}}\hat{K}_{1}( \zeta)  \right].
\end{align}

Substituting the explicit form of $N_\mu ^{\eff} $ and
simplifying, we obtain
\begin{align}
\label{nsplit4} & S_{\mu\nu}^{(-1)}  =  2{\kern 1pt}G b^{\kn 2}
e^2 \e^{i (kb)/2} \int\limits_0^1 dx \,\e^{-i (k b) x}
\hat{K}_{-1} \!\left( \zeta \right)
\left[\vp(1-x)(ku')+x(ku)\vph\right]^2 u_\mu u_\nu \,.
\end{align}

\textbf{4.} Consider, next $T^{I\akn I}_{\mu\nu}$. Using the
formulae derived in Appendix \ref{Local_ints}, i.e.
\begin{align}
\label{repres}
& \e^{-i (k b)} \Phi(b)=\frac{1}{4\pi} \int\limits_0^1 dx\;\e^{-i(kb)x}
 \left[x(2x-1)\, b^{\kn 2} k_{\perp}^2\hat{K}_{-1}(\zeta)+
2\left(\vph 1-i x (k b) \right) \!{K}_{0}( \zeta )\right] \nn \\
&   \Phi (b)=\frac{1}{4\pi} \int\limits_0^1 dx\;\e^{-i(kb)x}
 \left[(x-1)(2x-1)\, b^{\kn 2} k_{\perp}^2\hat{K}_{-1}(\zeta)+
2\left(\vph 1-i (x-1) (k b) \right) \!{K}_{0}( \zeta )\right],
\end{align}
$T^{I\akn I}_{\mu\nu}$ takes the form
\begin{align}
\label{bbb1}
 T^{I\akn I}_{\mu\nu}=- 2{\kern 1pt}G b^{\kn 2} e^2  \e^{i(kb)/2} \int\limits_0^1
dx\;\e^{-i(kb)x}&  \biggl[  \left[ \vph (x-1)(ku')^2
    +  x (ku)^2 \right] (2x-1)  \hat{K}_{-1}(\zeta)  + \\
&  \; \,+   2 \left[  (ku')^2\!\left(\vph 1-i (x-1) (k b) \right)
\! + (ku)^2\! \left(\vph 1-i x (k b) \right) \right]\!
\frac{{K}_{0}( \zeta )}{k_{\bot}^2 b^{\kn 2}} \biggr]
u_{\mu}u_{\nu}\,. \nn
\end{align}

\textbf{5.} Thus, the sum
\begin{align}
\label{bbb2} S_{\mu\nu}^{(-1)} + T^{I\akn I}_{\mu\nu} =- 2{\kern
1pt}G b^{\kn 2} e^2 \, \e^{i(kb)/2} \int\limits_0^1
dx\;\e^{-i(kb)x}& \biggl[ 2 \left[ (ku')^2\!\left(\vph 1-i (x-1)
(k b) \right) \! + (ku)^2 \!\left(\vph 1-i x (k b) \right) \right]
\! \frac{{K}_{0}(
\zeta )}{k_{\bot}^2 b^{\kn 2}}  - \nn   \\
& \,- \left[\vp(ku')+ (ku)\vph\right]^2 x (1-x)\,
\hat{K}_{-1}(\zeta) \biggr] u_{\mu}u_{\nu}\,,
\end{align}
and, using the identity $z^2\hat{K}_{-1}(z) =\hat{K}_{1}(z)$,
we obtain the explicitly finite expression
\begin{align}
\label{bbb3} S_{\mu\nu}^{(-1)} + T^{I\akn I}_{\mu\nu}=  \frac{2
{\kern 1pt} G e^2 }{k_{\bot}^2  }\, \e^{i(kb)/2}&  u_{\mu}u_{\nu}
\int\limits_0^1
dx\, \e^{-i(kb)x}  \biggl[    \left[\vp(ku')+ (ku)\vph\right]^2 \! \hat{K}_{1} (\zeta)  -\\
&  - 2 \left[ (ku')^2\!\left(\vph 1-i (x-1) (k b) \right) +
(ku)^2\! \left(\vph 1-i x (k b) \right) \right] {K}_{0}( \zeta )
\biggr] \nn
\end{align}
with no $\hat{K}_{-1}(\zeta) $.
{\it All divergent and $r_0-$dependent terms have cancelled.}

\subsection{The total amplitude}

Therefore the total effective radiation amplitude reads
\begin{align}
\label{bbb4}
   \tau_{\mu\nu} = S_{\mu\nu}^{{\kern 1pt}(0,1)} + S_{\mu\nu}^{(-1)} + T^{I\akn I}_{\mu\nu} +
   T_{\mu\nu}^{I\!I\!I}.
\end{align}

$\bullet$ From its definition in (\ref{nsplit3}), $ S_{\mu\nu}^{{\kern 1pt}(0,1)}$ takes the form
\begin{align}
S_{\mu\nu}^{{\kern 1pt}(0,1)} =8{\kern 1pt} G e^2 \, \e^{i
(kb)/2} \int\limits_0^1 dx \,\e^{-i (k b) x} & \left[
 - u_{\mu}u_{\nu}  {K}_{0}(
\zeta )- \frac{b_{\mu}b_{\nu}}{b^{\kn 2}}\hat{K}_{1}( \zeta )+ i
\left[\vp(1-x)(ku')+x(ku)\vph\right]
 u_{(\mu } b_{\nu )}  {K}_{0}\! \left(\zeta  \right) \right]. \nn
\end{align}

$\bullet$ Using formulae
\begin{align}
\label{repres2} & \e^{-i (k b)} \frac{\Phi'(b)}{b}=-\frac{ 1}{4\pi
b^{\kn 2}} \int\limits_0^1 dx\;\e^{-i(kb)x}
 \left[x(2x-1)\, b^{\kn 2} k_{\perp}^2\hat{K}_{0}(\zeta)+
2\left(\vph 1-i x (k b) \right) \!\hat{K}_{1}( \zeta )\right] \nn \\
&   \frac{\Phi'(b)}{b}=-\frac{ 1}{4\pi b^{\kn 2}} \int\limits_0^1
dx\;\e^{-i(kb)x}
 \left[(x-1)(2x-1)\, b^{\kn 2} k_{\perp}^2{K}_{0}(\zeta)+
2\left(\vph 1-i (x-1) (k b) \right) \!\hat{K}_{1}( \zeta )\right],
\end{align}
$\un T^{I\!I\!I}_{\mu\nu}$ in (\ref{TI-II-III}) is also written as
a sum of two integrals over $x$, one containing $K_0(\zeta)$ and
the other $\hat K_{1}(\zeta)$.

$\bullet$ Collecting terms with integrand proportional to $K_0$ and $\hat K_1$ we write  the total energy-momentum source $\tau_{\mu\nu}$ in the form
\begin{align}
\label{tau-mn}
 \tau_{\mu\nu}(k) =2{\kern 1pt}G e^2 \, \e^{i
(kb)/2} \int\limits_0^1 dx \,\e^{-i (k b) x}   \left[\,
Y_{\mu\nu}^{0}(k,x) \, {K}_{0}( \zeta )+ Y_{\mu\nu}^{1}(k,x)\,
\hat{K}_{1}( \zeta ) \right],
\end{align}
where
\begin{align}
\label{Y0Y1}
Y_{\mu\nu}^{0}=& -\left(4+\frac{2}{k_{\bot}^2} \left[\vp(ku')^2+
(ku)^2\vph\right]  + 4i \frac{(kb){\kern 1pt}x(1-x)}{k_{\bot}^2}
\left[\vp(ku')^2- (ku)^2\vph\right]\right)u_{\mu}u_{\nu}+ \nn
\\&+
8ix(1-x) \left[\vp(ku')+ (ku)\vph\right] u_{(\mu } b_{\nu )} \nn \\
Y_{\mu\nu}^{1}=& \left( \frac{1}{k_{\bot}^2}\left[\vp(ku')+
(ku)\vph\right]^2 + 4i \frac{(kb)}{k_{\bot}^4 b^{\kn 2}}
\left[\vp(ku')^2-(ku)^2\vph\right]- 4  \frac{(kb)^2}{k_{\bot}^4
b^{\kn 2}}
\left[\vp(1-x)(ku')^2+x(ku)^2\vph\right]\right)u_{\mu}u_{\nu} \nn
\\ &- 4 \frac{b_{\mu}b_{\nu}}{b^{\kn 2}}-\frac{8}{k_{\bot}^2 b^{\kn 2}  }\left( i\left[\vp(ku') +
(ku) \vph\right] + (kb)\left[\vp(x-1)(ku') +x(ku) \vph\right]
\right)u_{(\mu } b_{\nu )}\,.
\end{align}

$\bullet$ Defining in the center-of-mass frame the radiation
wave-vector by
\begin{equation} \label{kmu}
k^{{\kern 1pt}\mu} = \omega{\kern 1pt} (1, {\bf n}) = \omega{\kern
1pt} (1,\sin \vartheta \cos\varphi,\sin \vartheta \sin\varphi,
\cos \vartheta )
\end{equation}
and contracting \eqref{tau-mn} with the two polarizations (see
Appendix \ref{conventions}), we obtain the final (finite)
expressions for the source of the gravitational radiation
separately for the two polarizations\kn\footnote{Up to an overall
phase $e^{i (kb)/2}$ in both $\tau_+$ and $\tau_\times$, since it
does not contribute to the energy. }
\begin{align}
\label{tau+}
\tau_{+}(k) \equiv \tau_{\mu\nu}(k)\, \ep_+^{\mu\nu}=\frac{16 {\kern 1pt}G  e^2}{\sqrt{2} }\,
\int\limits_0^1 dx \,\e^{-i (k b) x}   \left[ - {K}_{0}(
\zeta )+\sin^2\!{\varphi}\, \hat{K}_{1}( \zeta )
\right],
\end{align}
and
\begin{align}
\label{taux}
 \tau_{\times}(k)\equiv \tau_{\mu\nu}(k)\, \ep_\times^{\mu\nu} =-\frac{16{\kern 1pt} G e^2}{\sqrt{2}}\, \sin \varphi\,
\!\!\int\limits_0^1 dx \,\e^{-i (k b) x}  \biggl[ 2i
\frac{\hat{K}_{2}( \zeta )-\hat{K}_{1}( \zeta )}{ \omega b \sin
\vartheta}  +  (2x-1)  \cos   \varphi  \,\hat{K}_{1}( \zeta )
\biggr]\,,
\end{align}
where $\zeta=\omega b \sin\vartheta \sqrt{x(1-x)}$.

{\it To summarize: The only approximation made so far is the restriction to the first order corrections of the gravitational field.  The leading non-linear terms were taken into account. To this order, the total source $\tau_{\mu\nu}$ (\ref{tau-mn}) and the separate sources of the two polarizations (\ref{tau+}) and (\ref{taux}) have been expressed as finite integrals over a parameter $x\in [0,1]$.}

\section{Characteristics of the emitted radiation}

We turn next to the computation of the emitted radiation frequency spectrum and of the total emitted energy. They are obtained from
 \begin{align}
 \label{fr_di}
\frac{d E_{\rm rad}}{d\omega \,d\Omega}=\frac{ G }{2\pi^{2}}
\,\omega^{2} \sum_{\cp} | \tau_{\cp}|^2\,,
\end{align}
summed over the two polarizations.

\begin{figure}
 \begin{center}
\includegraphics[
width=11cm]{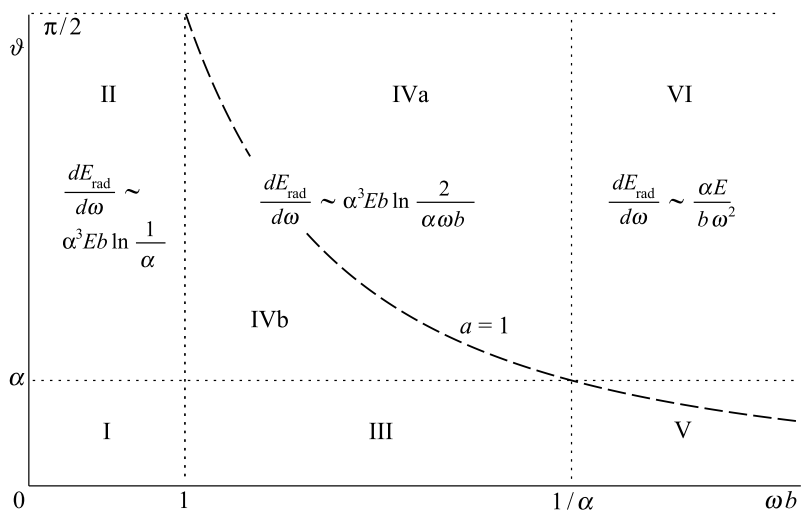}  \caption{The characteristic angular and frequency regimes.}
\label{table}
 \end{center}
\end{figure}

It will be convenient in the sequel to treat separately the six
angular and frequency regimes shown in Fig.\,\ref{table}.

\subsection{Zero-frequency limit  -- Regimes I and II} \label{I+II}

In the low-frequency regime ($\omega \to 0$) the amplitude $\tau_{\times} $
dominates and has the form
\begin{align}
\label{scalampl2dd}
 \tau_{\times} \simeq - \frac{16 \sqrt{2} i{\kern 1pt} G E^2 \, \sin \varphi}{\omega b \sin \vartheta}\,
\!\!\int\limits_0^1 dx \,\e^{-i (k b) x}  \left[\hat{K}_{2}( \zeta
)-\hat{K}_{1}( \zeta ) \right],
\end{align}
while $\tau_{+}$ is finite and gives subleading contribution to
\eqref{fr_di}. Note that in this limit ($\hat{K}_{1}( 0)=1$,
$\hat{K}_{2}(0)=2$, $\e^{-i (kb) x}=1$) the $x-$integration is
trivial and gives $d E_{\rm rad }/d\omega = (2^8  G^3 E^4/ \pi
b^{\kn 2}) \int d\vartheta/\sin\vartheta$, which diverges and
implies that our formulae are not valid for $\vartheta$ close to
zero.

We cannot trust our formulae in that regime and should repair
them. A quick way to do it, is to impose a small-angle cut-off
$\vartheta=\vartheta_{\rm cr}$ on the $\vartheta-$integration, so
as to obtain for $dE_{\rm rad}/d\omega|_{\omega=0}$ the value
computed quantum mechanically in \cite{Weinberg,Wbook},
namely
\begin{align}
\label{DEMD_ef1}
\left( \frac{dE^W}{d\omega} \right)_{\!\omega=0}& = \frac{4}{
 \pi} \,  G  |t| \ln \frac{s}{|t|}=  \frac{G   \alpha^2 s}{
 \pi}\,   \ln \frac{4}{ \alpha^2} = \frac{E\kn b}{\pi}\, \alpha^3 \ln(2/\alpha) \,, \qquad s=4E^2\,,
\end{align}
which indeed agrees with our expression
 \begin{align}
 \label{fr_di2}
\left( \frac{dE_{\rm rad}}{d\omega} \right)_{\!\omega=0}=\frac{
2^9 G^3 E^4 }{ \pi b^{\kn 2} }
 \int\limits_{\thcr}^{\pi/2} \frac{d\vartheta}{\sin \vartheta}= \frac{E\kn b}{\pi}\, \alpha^3 \,\ln  \ctg \frac{\thcr}{2}
\end{align}
for
\begin{align}
\label{thetacr}
\thcr =\alpha \ll 1\,.
\end{align}

\bigskip

\textit{Thus, our result for the low frequency radiation emitted
in a collision with large $s$, fixed $t=-(s/2)\, (1-\cos\alpha)
\simeq -s\alpha^2/4$ and $G|t| \leqslant {\mathcal O}(1)$, agrees
with the quantum computation of Weinberg, apart from a tiny
emission angle \mbox{$\vartheta \leqslant {\mathcal O}(\alpha) \ll
1$} in the forward direction\kn\footnote{It should be pointed out
that our classical computation reproduces the quantum results of
Weinberg for soft graviton emission for $G|t| \ll {\mathcal
O}(1)$, i.e. for $(E/M_{\rm Pl})(r_S/b)\ll 1$, or in Weinberg's
notation for $B \ll 1$.}.   Furthermore, as will be shown next, at
low frequencies $\omega$ and for \mbox{$\vartheta \geqslant \alpha
$}, the leading contribution to our classical amplitude, dominated
in this regime by $T^{I\! I\!I}_{\mu\nu}$ (\ref{TI-II-III}), is
identical to the one obtained in \cite{Wbook}, after it is
generalized (see below) to $b\neq 0$\kn\footnote {It is not
surprising that the quantum and the classical results agree for
the emitted energy of low frequency. As we argued, in the
low-frequency regime, $dE_{\rm rad}/d\omega$ is dominated by the
local source. The contribution of stress in this regime is
negligible. As a result, the radiated gravitons are expected to be
produced in a coherent state \cite{Glauber}, and the corresponding
expectation value of the quantum field to satisfy the classical
field equations.}.}

Indeed, following the notation of \cite{Wbook}, we write for the
energy-momentum source of the $2 \to 2$ scattering process we are
studying for arbitrary, a priori, scattering angle $\alpha$ and
impact parameter $b$:
\begin{align}
& \tilde{T}^{\mu\nu}({\bf x},t)=\sum_{n=1}^2 \frac{P_n^\mu
P_n^\nu}{E_n}\, \delta^3({\bf x}- {\bf v}_n t \mp {\bf b}/2)\,
\theta(-t) +
\sum_{n=1}^2 \frac{\tilde P_n^\mu \tilde P_n^\nu}{\tilde E_n} \,\delta^3({\bf x}- \tilde{\bf v}_n t \mp {\bf b}/2) \,\theta(t) \nn \\
&=\sum_{n=1}^2 \frac{P_n^\mu P_n^\nu}{E_n} \, \delta^3({\bf x}-
{\bf v}_n t \mp {\bf b}/2) + \sum_{n=1}^2 \biggl(\frac{\tilde
P_n^\mu \tilde P_n^\nu}{\tilde E_n} \,\delta^3({\bf x}- \tilde{\bf
v}_n t \mp {\bf b}/2) - \frac{P_n^\mu P_n^\nu}{E_n}
\,\delta^3({\bf x}- {\bf v}_n t \mp {\bf b}/2)\biggr) \theta(t)
\nonumber
\end{align}
where $P_n^\mu (\tilde P_n^\mu), n=1,2$, are the initial (final) particle momenta, with $\tilde P_n=\tilde P_n (P_n, \alpha )$.
Its Fourier transform is
\begin{align}
\tilde T^{\mu\nu}({\bf k}, \omega)&=\sum_{n=1}^2 2\pi  \e^{\pm
i(kb)/2} {P_n^\mu P_n^\nu}\, \delta( k\cdot P_n) + \sum_{n=1}^2
\e^{\pm i(kb)/2}\biggl(\frac{\tilde P_n^\mu \tilde P_n^\nu}{\tilde
E_n} \frac{i}{\omega-{\bf k}\tilde{\bf  v}_n} - \frac{P_n^\mu
P_n^\nu}{E_n} \frac{i}{\omega-{\bf k}{\bf v}_n} \biggr), \nn
\end{align}
where $k^\mu = (\omega, {\bf k})$ is the radiation wave-vector.
The terms \mbox{$n=1$} (\mbox{$n=2$}) in the sums, are multiplied
by $\e^{+i(kb)/2}$ $(\e^{-i(kb)/2})$, respectively.

The first sum, proportional to delta-functions, corresponds to no scattering and does not
contribute to radiation. Thus, we end-up effectively with
\begin{align}
\label{GWM_Fourier1} \tilde T^{\mu\nu}_{\eff}(k)&=  i\sum_{n=1}^2
\e^{\pm i(kb)/2}\biggl(\frac{\tilde P_n^\mu \tilde P_n^\nu}{\tilde
E_n} \frac{1}{\omega-{\bf k} \tilde{\bf v}_n} - \frac{P_n^\mu
P_n^\nu}{E_n} \frac{1}{\omega-{\bf k}{\bf v}_n} \biggr).
\end{align}

To leading order in our approximation the scattering process, we
are dealing with, is elastic with $\tilde{E}_n=E_n=E$.
Furthermore, write for the incoming particles $P_n^\mu = E u_n^\mu
= E (1,0,0,\pm1)$ and for the outgoing ones $\tilde P_n^\mu=E
\tilde u_n^\mu = E (u_n^\mu + \!\un \dot z_n^\mu)=E(u_n^\mu \mp
\alpha\, \hat b^{\kn\mu}) \equiv P_n^\mu +\! \un P_n^\mu$,
substitute into $\tilde T_{\rm eff}^{\mu\nu}$, and expand in
powers of $\alpha$ using the fact that for $\vartheta > \alpha$
one has $|k \cdot\! \un P_n^\mu| \ll |k  \cdot \! P_n^\mu|$, to
obtain
\begin{align}
\label{GWM_Fourier3} \tilde T^{\mu\nu}_{\eff}(k) = i \sum_{n=1}^2
\e^{\pm i(kb)/2}\biggl(\frac{   2   P_n^{(\mu}  \un P_n^{\nu)}}{
(k P_n) } - \frac{ (k\un P_n)}{ (k P_n)^2 }\, P_n^{\mu}P_n^{\nu}
\biggr) +\mathcal{O}(\alpha^2)
\end{align}
or, finally, making use of the above definitions,
\begin{align}
\label{GWM_Fourier5} \tilde T^{\mu\nu}_{\eff}(k) = \frac{i \alpha
E}{b} \sum_{n=1}^2 \pm\, \e^{\pm i(kb)/2}\frac{  (kb)
\,u_n^{\mu}u_n^{\nu}-  2 (k u_n)\, u_n^{(\mu}
b^{\nu)}_{\vphantom{n}}}{ (k u_n)^2 }=  \frac{8 i G    E^2
}{b^{\kn 2}} \left[\e^{i(kb)/2}
 \frac{ \sigma^{(u)}_{\mu\nu}}{ (ku)^2 } - \e^{-i(kb)/2} \frac{  {\sigma}^{(u')}_{\mu\nu}}{ (ku')^2 }  \right] \,.
\end{align}
This is {\it identical} to $T^{I\!I\!I}_{\mu\nu}$ in
(\ref{TI-II-III}), after we bring the latter to its original form
(\ref{gg1D'}) by the substitution $\bar{\sigma}^{(u)}_{\mu\nu}\to
\sigma^{(u')}_{\mu\nu}$. Q.E.D.

{\it Summary: In regime II we use our formula, which is identical
to Weinberg's. We extend the use of Weinberg's formula in regime I
as well. The total emitted energy in I is known to be of
$\,{\mathcal O}(\alpha^3 E)$. Given that $(dE_{\rm
rad}/d\omega)_{\omega=0}\sim (dE_{\rm rad}/d\omega)_{\omega=1/b}$,
the contribution of II to the radiation efficiency is estimated by
multiplying (\ref{fr_di2}) by the frequency range $1/b$ and
dividing by the initial energy $2E$. The result is}
\begin{equation} \epsilon_{I,I\akn I} \sim \alpha^3 \ln(1/\alpha)\,.
\end{equation}

\subsection{Regime VI}

Here we consider the Regime VI ($\omega> 1/(\alpha b) =1/2r_S$ and
$ \vartheta> \alpha$). Contrary to regimes I and II, in regime VI
as well as in IV, which will be discussed in the next subsection,
the contributions to radiation of the local term and the stress
are equally important.

The cross-amplitude (\ref{taux}) after an integration by parts of the second term reads:
\begin{align}
 \tau_{\times}(k)=-\frac{32i{\kern 1pt} G e^2}{\sqrt{2}}\, \sin
 \varphi
 \left[\,\int\limits_0^1 dx \,\e^{-i (k b) x}  \! \left(
\frac{\hat{K}_{2}( \zeta ) \sin^2\!   \varphi-\hat{K}_{1}( \zeta
)}{a}    \right) - \frac{4\cos   \varphi}{a^2}\,\e^{-i (kb)/2}
\sin \frac{kb}{2} \right]\nn
\end{align}
with $a\equiv \omega b \sin\vartheta$ and together the two amplitudes are written as
\begin{align}
& \tau_{+}(k)=\frac{16 {\kern 1pt}G  e^2}{\sqrt{2} }\, \left[ -
{L}_{0}+ {L}_{1} \sin^2\!{\varphi}\,  \vph\right], \nn \\ &
 \tau_{\times}(k)=-\frac{32i{\kern 1pt} G e^2}{\sqrt{2}}\, \sin \varphi\,
 \left[
\frac{L_{2} \sin^2\!   \varphi  -L_{1} }{a}  - \frac{4\cos
\varphi}{a^2}\,\e^{-i (kb)/2} \sin \frac{kb}{2} \right],
\end{align}
where
$$L_{m}(a,\varphi)\equiv \int\limits_0^1 \e^{iax\cos \varphi} \hat{K}_m \!\left(a
\sqrt{x(1-x)}\right)dx$$ is defined and studied in Appendix
\ref{Lm}.

Substituting the large-$a$ expansion  to  leading order
$$ L_m \simeq
\frac{2^{m+2}\Gamma(m+1)}{a^2}\,\,\e^{ia\cos
\varphi/2}\cos\frac{a\cos \varphi}{2}\,,
 $$
we obtain
\begin{align}
& \tau_{+}(k)\approx -\frac{64 {\kern 1pt}G  e^2}{\sqrt{2} a^2}\,
 \cos 2 {\varphi}  \cos \frac{a\cos \varphi}{2}\,, \qquad
 \tau_{\times}(k) \approx  -\frac{64 i{\kern 1pt} G e^2}{\sqrt{2} a^2}\,
 \sin 2
 \varphi\, \sin
 \frac{a\cos \varphi}{2}\,.
\end{align}

Thus
\begin{align}
\frac{dE_{\rm rad}}{d\omega d\Omega}= \frac{2^{10} G^3 E^4 }{
\pi^{2}}  \frac{\omega^{2}}{a^4}\biggl[  \cos^2\! 2 {\varphi}
\cos^2\!\left(\frac{a\cos \varphi}{2}\right) +\sin^2 \!2
 \varphi\sin^2 \!\left(\frac{a\cos \varphi}{2}\right) \biggr].
\end{align}
Integrate over $\varphi$ using the formulae (\ref{integrals})
to obtain for $a\gtrsim {\mathcal O}(1)$
\begin{align}\label{dwdth}
\frac{dE_{\rm rad}}{d\omega d \vartheta}= \frac{2^{10} G^3 E^4 }{
\pi } \frac{\omega^{2}}{a^4}\left[ 1-6\frac{J_1(a)}{a}- \left(1 -
\frac{24}{a^2}\right)
  J_2(
  a)\right]\sin \vartheta \sim \frac{2 (\alpha b)^3 E }{\pi}\, \frac{\omega^{2}}{a^4}
  \sin\vartheta\,,
\end{align}
from which one can obtain an estimate for the frequency distribution of the emitted radiation in regime VI by integrating over $\vartheta \in (\alpha, \pi-\alpha)$, namely
\begin{align}\label{dw}
\frac{dE^{V\!I}_{\rm rad}}{d\omega }  \sim \frac{\alpha E }{
b} \,\frac{1}{\omega^{2} } \,, \qquad \omega>
1/\alpha b\,,
\end{align}
as well as an estimate for the emitted energy and the corresponding efficiency in regime VI, by integrating also over $\omega\in (1/\alpha b, \infty)$,
\begin{align}
\label{Erad} E^{V\!I}_{\rm rad}\sim \alpha^2 E  \qquad {\rm and}
 \qquad \epsilon_{V\!I} \sim \alpha^2.
\end{align}

\subsection{Regimes II+IV+VI}

Finally, we shall discuss the characteristics of the radiation
corresponding to the union of regimes  II+IV+VI. For that, it is
convenient to transform the amplitudes (\ref{tau+}) and
(\ref{taux}) to the form
\begin{align} \label{scalampl}  & \tau_{+} (a, \varphi)=\frac{16 {\kern 1pt}G  e e'}{\sqrt{2} }\,
  \int\limits_0^1 dx \,\e^{i a x\cos
\varphi} \left[ - {K}_{0}( \zeta )+\sin^2\!{\varphi}\,
\hat{K}_{1}( \zeta )
\right], \\
& \tau_{\times}(a, \varphi) =-\frac{16i{\kern 1pt} G e
e'}{\sqrt{2}}\, \sin \varphi  \!\int\limits_0^1 dx \,\e^{i a x\cos
\varphi} \! \left[2  x{\kern 1pt}(1-x){\kern 1pt}a {\kern 1pt}
{K}_{0}( \zeta )+ \left(-i (2x-1) \cos \varphi +\frac{2 }{ a}
  \right) \!\hat{K}_{1}( \zeta ) \right].\nn
\end{align}


Representing Macdonalds as
\begin{align} \label{cher1}
{K}_{0}( \zeta )=\int\limits_0^{\infty}  \frac{\exp \left(- \zeta
\sqrt{t^2+1} \right)}{\sqrt{t^2+1}} \, dt\,,  \qquad\qquad
\hat{K}_{1}( \zeta ) = \zeta^2 \int\limits_0^{\infty} \frac{\exp
\left(- \zeta \sqrt{t^2+1} \right)}{\sqrt{t^2+1}}\,t^2\, dt\,,
\end{align}
we square and sum up the two polarizations:
\begin{align} \label{cher2}
 & |\tau_{+} (a, \varphi)|^2 +  |\tau_{\times} (a, \varphi)|^2= 2^7 {\kern 1pt}(G  e^2)^2 \,
  \int\limits_0^1 dx  \int\limits_0^1 dx' \int\limits_0^{\infty} dt  \int\limits_0^{\infty} dt'
  \,\frac{\e^{-
  a  \Lambda(x,\kn t,\kn x'\!,\kn t'\!,\kn \varphi)}}{\sqrt{t^2+1}\sqrt{t'^2+1}  } \, \sum_{k=0}^4
M_k(x,t,x'\!,t'\!, \varphi)\,a^k,\nn \\
& \Lambda(x,t,x'\!,t'\!, \varphi) = \sqrt{x(1-x)} \sqrt{t^2+1}+
\sqrt{x'(1-x')} \sqrt{t'^2+1} -i (x-x') \cos \varphi
\end{align}
with
\begin{align} \label{cher3}
 & M_0=1 \nn \\
 & M_1=0 \nn \\
 & M_2=  x(1-x) \sin^2\! \varphi\left[ -   t^2 +2
x' (1-x') (1+2\kn t^2+t^2 t'^2) \vph\right]+\{  x
\longleftrightarrow x', t \longleftrightarrow t'\} \nn \\
  & M_3= 2 i
xx'(1-x)(1-x')\,t'^2 \sin^2\!\varphi\cos \varphi \left[2x'-  1+  2
x't^2  \vph \right]-\{  x
\longleftrightarrow x', t \longleftrightarrow t'\}\nn \\
 & M_4=  x x' (1-x) (1-x')\, t^2 t'^2\sin^2\!\varphi\left[\sin^2\!\varphi +(2x-1)(2x'-1) \cos^2\!
 \varphi  \vph\right].
\end{align}

\begin{figure}
\begin{center}
\noindent
\subfigure[]{\raisebox{0pt}{\includegraphics[width=0.49\textwidth]{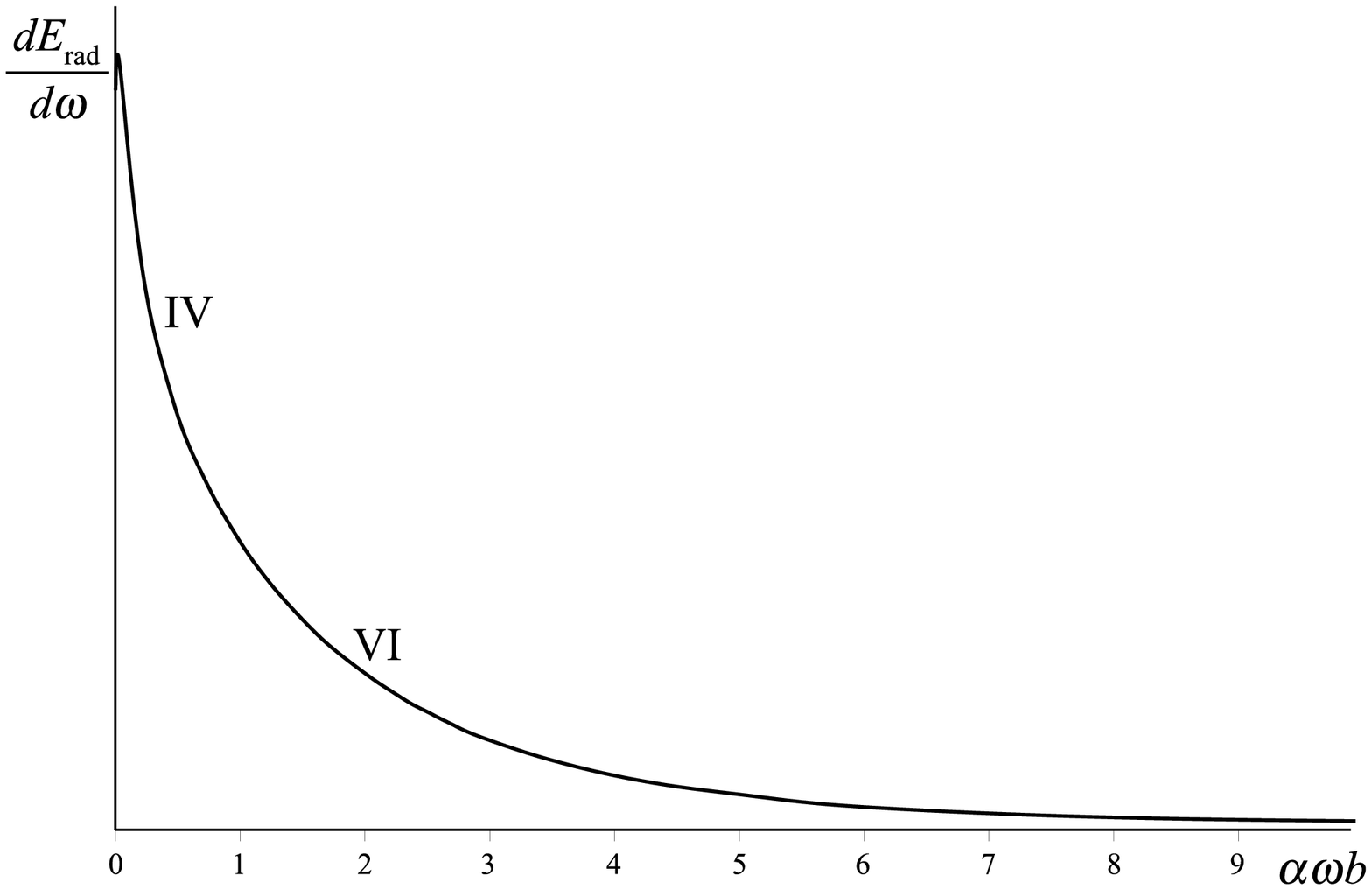}\label{w-dist}}}\qquad
\quad
\subfigure[]{\raisebox{0pt}{\includegraphics[width=0.45\textwidth]{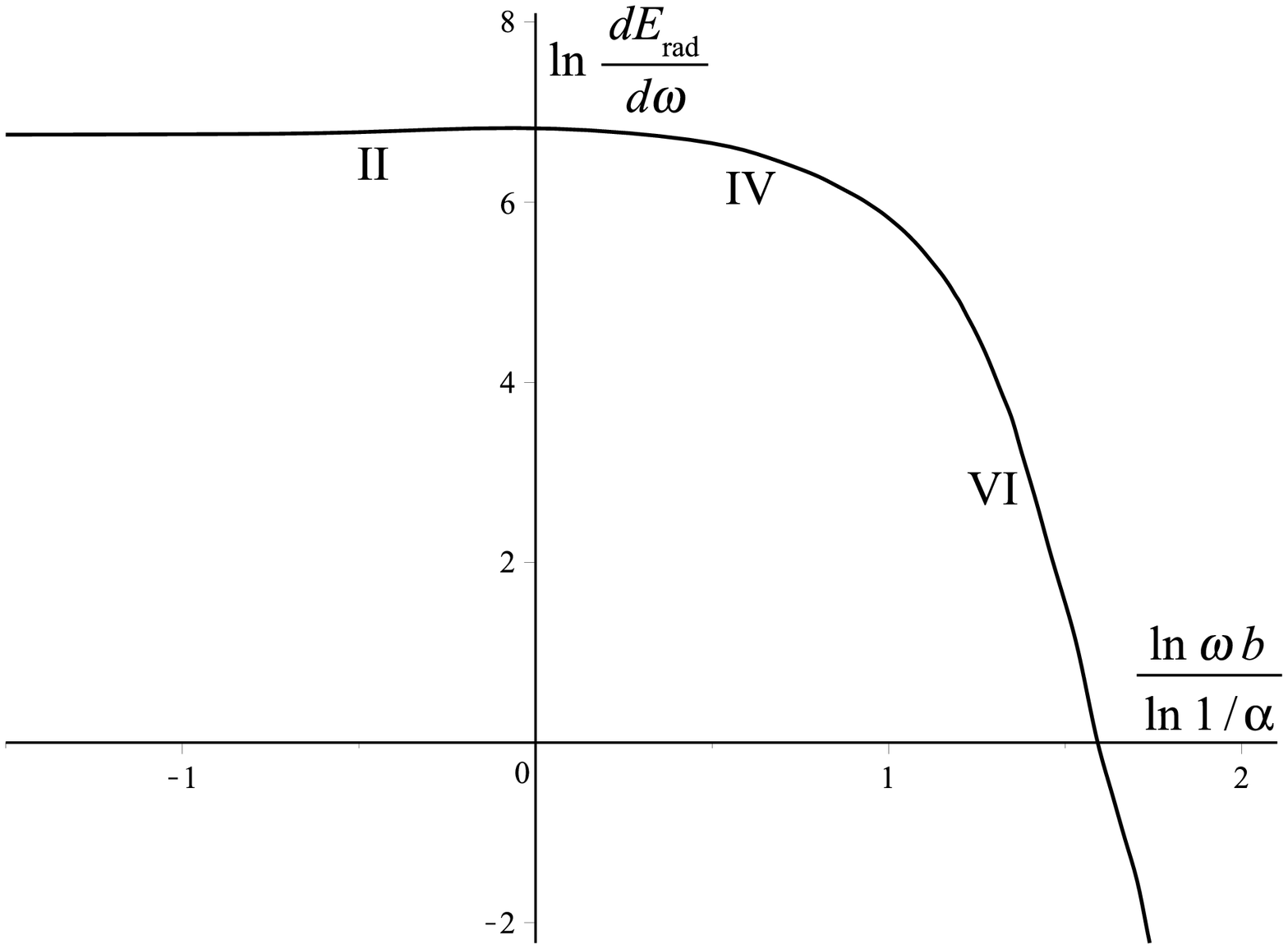}\label{log-log}}}
\caption{The frequency distribution in the combined regimes
II+IV+VI for $\alpha=0.01$ with two different choices of axes'
labelings. The regime II in the Figure on the left is compressed
in the interval $(0, \alpha)$. The slope in regime VI of the right
plot corresponds to $dE_{\rm rad}/d\omega \sim 1/\omega^2$.}
\label{FigII+IV+VI}
\end{center}
\end{figure}

One may substitute into (\ref{fr_di}) and integrate numerically
over all variables apart from $\omega$ to obtain $dE_{\rm
rad}/d\omega$. The result is shown in Fig.\,\ref{FigII+IV+VI}. On
the other hand, one may change variables and integrate over $a$
instead of $\omega$ to obtain:
\begin{align}
 \frac{dE_{\rm rad}}{d \vartheta}=\frac{2^6 {\kern 1pt}G^3  E^4}{
 \pi^2 b^{\kn 3}} \frac{1}{\sin^2\!\vartheta}  \sum_{k=0}^4 (k+2)!
 \int\limits_0^1 dx \int\limits_0^1 dx'
\int\limits_0^{\infty} \frac{dt}{\sqrt{t^2+1}}
 \int\limits_0^{\infty} \frac{dt'}{\sqrt{t'^2+1}}\int\limits_0^{2\pi} d\varphi\,\frac{ M_k
 }{\Lambda^{k+3} }\,.\nn
\end{align}
The convergence of this 5-variable integral follows from the
behavior of the integrand at large-$a$, while its
numerical value is expected to be of $\mathcal{O}(1)$, since the integrand does not contain any  small or large parameter. Indeed, numerical
integration (including the factorial in front) leads to
\begin{align} \label{cher5}
  \eta \equiv   \sum_{k=0}^4 (k+2)!
 \int\limits_0^1 dx \int\limits_0^1 dx'
\int\limits_0^{\infty} dt
 \int\limits_0^{\infty} dt'\int\limits_0^{2\pi} d\varphi\frac{ M_k
 }{\Lambda^{k+3}\sqrt{t^2+1}\sqrt{t'^2+1} } \approx 89.9\,.
\end{align}

Thus the angular distribution reads
\begin{align} \label{cher6}
 \frac{dE_{\rm rad}}{d \vartheta}=\frac{\eta \alpha^3 E}{8\pi^2}
 \frac{1}{\sin^2\!\vartheta}\,.
\end{align}

Integration over $ \vartheta\in (\alpha,\pi-\alpha)$ gives
\begin{align} \label{cher7}
 E_{\rm rad} =\frac{\eta {\kern 1pt}\alpha^2 E}{4\pi^2}
\end{align}
and for the efficiency
\begin{align} \label{cher8}
\epsilon =\frac{ E_{\rm rad}}{2E} \simeq 1.14\,\alpha^2\,.
\end{align}

 \begin{figure}
 \begin{center}
\includegraphics[
width=10cm]{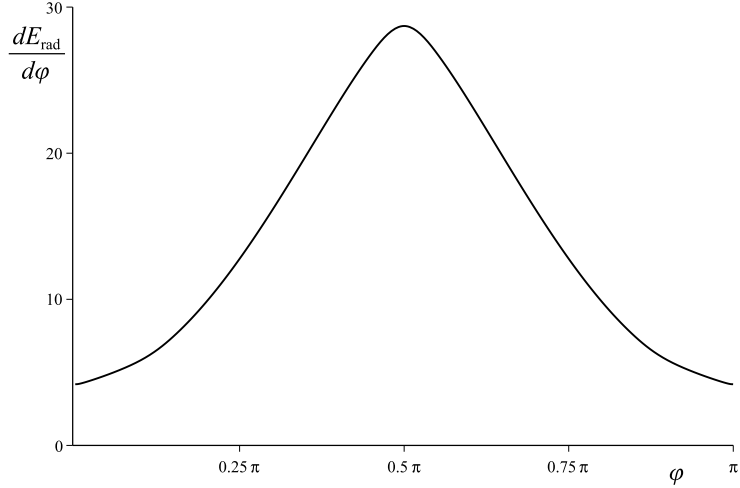}  \caption{The
$\varphi-$distribution for $G=b=1$ and $\alpha=0.01$.}
\label{phi-dist}
 \end{center}
\end{figure}

Finally, the $\varphi-$distribution of the emitted radiation is
shown in Fig.\,\ref{phi-dist}, according to which most energy is
emitted perpendicular to the scattering plane.

\bigskip

It is instructive to study regime IV in a little more detail. For
that, let us split regime IV into IVa and IVb, as shown in
Fig.\,\ref{table}, according to \mbox{$a>1$} and \mbox{$a<1$},
respectively. Inside the regime IVa the amplitude is damped as in
regime VI. However, near the left border of regime IVb (with
$1/b\lesssim \omega \ll 1/\alpha b$) one may expand the amplitudes
in powers of $a$ and obtain, as in regime II:
\begin{align}
\label{ZFL2k}
\tau \simeq \tau_{\times} \simeq
-\frac{16 \sqrt{2}  i   G E^2}{ \omega b} \frac{\sin \varphi}{\sin
\vartheta}\,.
\end{align}
Upon integration over regime IVb, i.e. for $\alpha \lesssim \vartheta \lesssim \vartheta_{\rm max}=\arcsin(1/\omega b)$ one obtains
 \begin{align}
\left( \frac{dE_{\rm rad}}{d\omega} \right)_{\!1/b\lesssim
\omega\ll 1/\alpha b}\simeq
 \frac{\alpha^3  E\kn b}{\pi} \,\ln \frac{ \tg(\vartheta_{\max} /2)}{\tg(\alpha/2)}
\simeq \frac{\alpha^3  E \kn b}{\pi}\,\ln \frac{2 \alpha^{-1}
}{\omega b + \sqrt{\omega^2 b^{\kn 2} -1}}\,. \nn
\end{align}
Thus, for $1/b\lesssim \omega \ll 1/\alpha b$ one may approximate
$dE_{\rm rad}/d\omega$ by
\begin{equation}
\left( \frac{dE_{\rm rad}}{d\omega} \right)_{\!1/b\lesssim
\omega\ll 1/\alpha b}\simeq \frac{\alpha^3  E \kn b}{\pi}\,\ln
\frac{2}{\alpha \omega b}\,. \nn
\end{equation}
On the other hand, from the known behavior near $1/\alpha b\,$ from
regime VI, we know that
\begin{equation}
\left( \frac{dE_{\rm rad}}{d\omega} \right)_{\omega\sim 1/\alpha
b}\sim {\alpha^3  E\kn  b}\,. \nn
\end{equation}
So, a natural interpolation of $dE_{\rm rad}/d\omega$ between the
values $1/b$ and $1/\alpha b$ is
\begin{equation} \label{fr_di2m}
\left(\frac{dE_{\rm rad}}{d\omega} \right)_{\!1/b\lesssim
\omega\lesssim 1/\alpha b} = \frac{\alpha^3 E \kn b}{
\pi}\,\xi(\omega b) \,\ln \frac{2}{\alpha\omega b}\,.
\end{equation}
It may be shown numerically that $\xi(\omega b)$ is a slowly
varying function of ${\mathcal O}(1)$ in the regime $1/b\lesssim
\omega\lesssim 1/\alpha b$, so that one can simply write
instead
\begin{equation} \left( \frac{dE_{\rm rad}}{d\omega}
\right)_{\!1/b\lesssim \omega\lesssim 1/\alpha b} \sim  \alpha^3
E\kn b\, \ln \frac{2}{\alpha \omega b} \,.
\end{equation}

It should be pointed out here that the integral of $dE_{\rm
rad}/d\omega$ over $\omega$ receives most of its contribution from
frequencies in the neighborhood of $1/\alpha b$ in both regimes IV
and VI. Thus, one can say that the characteristic frequency of the
emitted radiation is around ${\mathcal O}(1/r_S)$.

\subsection{Comparison to previous work}

Let us briefly compare the results of the present paper with our
previous calculation of massive particle collisions and  to the
previous literature as well. In relation to our work, we should
point out that it is not straightforward to compare with
\cite{GST}, simply because they are were based on different
assumptions and approximations. In particular, in the massive case
we worked in the lab-frame and assumed weak-field approximation, restricting $\gamma_{\rm cm}$
to \mbox{$\gamma_{\rm cm} \ll 1/\alpha$}, which is not consistent
for fixed $\alpha$ with the \mbox{$\gamma_{\rm cm}\to \infty$}
limit, necessary to connect with the massless case. On the other hand, in the massless case the
computation is characterized by \mbox{$\gamma_{\rm cm}=\infty$},
is performed in the center-of-mass frame, used the scattering
angle $\alpha$ as an angular integration cut-off,  and used
Weinberg's results in regime I. So, before it is appropriate to
compare with the massive case, we have to redo the latter
following the same methodology and approximations. This, as well
as comparison with other papers in the literature is the content
of this subsection.

In \cite{GST} we computed the gravitational radiation efficiency
in ultra relativistic massive particle collisions in
$D$ dimensions and in the weak field limit, i.e. for \mbox{$1\ll
\gamma_{\rm cm} \ll 1/\alpha$}. Specifically, in $D=4$ we obtained
\mbox{$\epsilon\sim \alpha^3 \gamma_{\rm cm}$}, with the emitted
radiation having characteristic frequency \mbox{$\omega_{\rm
cm}\sim \gamma_{\rm cm}/b$} and emission angle
\mbox{$\vartheta\sim 1/\gamma_{\rm cm}$}, where $b$ is the impact
parameter. Thus, the efficiency is finite in the above kinematical
regime and agrees with the results of \cite{Peters} and
\cite{Thorne}.

However, to compare with the massless case we have to
improve the massive case computation in two ways: (i) discard the
weak-field condition as it is done here and more generally in the literature, e.g. \cite{ACV,GV,Dray87}, and (ii) use
the scattering angle $\alpha$ as a characteristic angle to cut-off
angular integrations. As a consequence of (i)
the condition \mbox{$1\ll \gamma_{\rm cm} \ll 1/\alpha$} is
replaced by \mbox{$\gamma_{\rm cm} \gg 1$} and \mbox{$1/\alpha\gg
1$}, independently; so that for \mbox{$\gamma_{\rm cm}\to \infty$}
one would naively conclude that $\epsilon \sim \alpha^3
\gamma_{\rm cm} \to \infty$, in disagreement with the present
massless result $\epsilon \sim \alpha^2$. But, as we will argue
next, the formula $\epsilon \sim \alpha^3 \gamma_{\rm cm}$ is not
valid in this case and the present result from regimes I-IV and
VI, defined by $\alpha$, $1/b$ and $1/\alpha b$ as in Fig.
\ref{table}, is actually in agreement with the limit
\mbox{$\gamma_{\rm cm}\to\infty$} of the contribution from those
regimes in the massive case.

Let us start with regime ${\rm IV_a+VI}$. Due to destructive
interference \cite{GST}, the total radiation amplitude
is
\begin{align}
\tau(\omega,\vartheta) \sim \frac{G m^2 \gamma^2_{\rm cm}}{(\omega
b \sin\vartheta)^2}\, ,
\end{align}
 which upon
integration over this region gives $\epsilon \sim \alpha^2 +
{\mathcal O}\kn(1/\gamma_{\rm cm})$\kn.

The contribution from regimes ${\rm II+IV_b}$ can be computed as
in the massless case. That is, starting with the observation that
in that regime the direct amplitude $T$ dominates over the stress
$S$, we take for $T$ the expression
\cite[Eqn.\,(3.17)]{GST}\footnote{In this expression the true
4-velocity $U^{\mu}$ normalized by $U^2=1$, is defined as
$U^{\mu}=\gamma_{\rm cm} u^{\mu}$ with $u^{\mu}_{1,2}=(1,0, 0,\pm
v_{\rm cm})$ and $v_{\rm cm}^2=1-\gamma_{\rm cm}^{-2}$.}
\begin{align}\label{massive}
\un T^{\mu\nu}(k)\sim \frac{i\kn G m^2 \gamma_{\rm cm}^2}{b^2}
\sum_{n=1}^2 \pm \,\e^{\pm i(kb)/2}\frac{  (kb)
\,u_n^{\mu}u_n^{\nu}-  2 (k u_n)\, u_n^{(\mu}
b^{\nu)}_{\vphantom{n}}}{ (k u_n)^2 }\,
 \hat{K}_{1}\!\left( \frac{(k u_n)\kn b}{2\gamma_{\rm cm} } \right)
\end{align}
 and integrate over the above regime to obtain again
$\epsilon \sim \alpha^2$, plus subleading corrections of order
$1/\gamma_{\rm cm}$\,.

Next, one can, following the approach discussed in subsection
\ref{I+II} above, use instead of (\ref{massive}) the expression
$\tilde T^{\mu\nu}_{\rm eff}$ given in (\ref{GWM_Fourier1}), valid
for both massive and massless colliding particles. This allows us
to integrate over all angles and upon integration over I+III to
obtain again to \mbox{$\epsilon \sim \alpha^2$}.  Furthermore,
this approach eliminates the disagreement of
\cite[Eqn.\,(4.15)]{GST} as well as of Smarr
\cite[Eqn.\,(2.13)]{Smarr} in the expression for $dE_{\rm
rad}/d\omega$ in the zero frequency limit with the corresponding
result of Weinberg. Specifically, the non-sensical for
$\gamma_{\rm cm} \to \infty$ formula $(dE_{\rm
rad}/d\omega)_{\omega=0}\simeq (G s\alpha^2/\pi)
\ln(4\gamma^2_{\rm cm})$  gets modified to $(dE_{\rm
rad}/d\omega)_{\omega=0}\simeq (G s\alpha^2/\pi)\ln (s/|t|)$ given
in (\ref{DEMD_ef1}).

Finally, in regime V the direct amplitude and the stress are
equally important, both for massive as well as massless colliding
particles. This was important in deriving destructive
interference, but on the other hand it does not allow us to
replace the source amplitude with Weinberg's formula for $T$. To
proceed along these lines, one has to ``guess" the correct
corresponding modification of $S$, consistent with the
conservation requirements for the total energy momentum tensor.
This problem is currently under investigation.

Thus, indeed, the results of the present paper can be obtained as
the massless limit of the corresponding conclusions of the massive
case in their common region of validity.

\section{Conclusions -- Discussion}

Using the same approach as in \cite{GST}, based on standard GR,
with the leading non-linear gravity effects taken into account, we
studied collisions of massless particles and computed the
gravitational energy of arbitrary frequency, which is emitted
outside the cone of angle $\alpha =2 r_S/b \ll 1$ in the forward
and backward directions. The value $\epsilon \simeq 1.14\,
\alpha^2$ was obtained for the radiation efficiency, with
characteristic frequency $\omega \sim 1/r_S$.  In fact, this value
represents a lower bound of the efficiency, since it does not
include the energy emitted inside that cone. The frequency
distribution of radiation in the characteristic angle-frequency
regimes is shown in Fig.\,\ref{table}. Our method allowed to study
the zero frequency regime and showed that the values $dE_{\rm
rad}/d\omega$ at $\omega=0$ and $1/b$ are of the same order.

We would like to point out that our results about regime IV agree
with Gruzinov and Veneziano \cite{GV}. Furthermore, our work
provides information about the very low frequency regime II, in
which, strictly speaking, the method of \cite{GV} cannot be
applied. Finally, at frequencies $\omega\lesssim1/\alpha b$ we
obtain $\epsilon \sim \alpha^2$, i.e. the same dependence as in
\cite{GV}. A possible difference in the coefficients is expected,
since  the  {\it non-linear} stress contribution around
$\omega\sim 1/\alpha b$ is comparable to the contribution of the
direct linearized emission from the colliding particles. However,
as mentioned several times, our approach cannot yet deal reliably
with regime V, which according to \cite{GV} gives dominant
contribution in $\epsilon$ by an extra factor $\ln 1/\alpha$. In
particular, we cannot yet confirm the presence of any other
characteristic frequency, such as e.g. $1/\alpha^3 b$, or
characteristic emission angle smaller than $\alpha$ \cite{GV}. We
hope to return to these issues with a better understanding of
regime V  in the near future.

\bigskip
{\bf Acknowledgments.} We would like to acknowledge enlightening
discussions with D.\,Gal'tsov, G.\,Veneziano, A.\,Gruzinov and
S.\,Dubovsky. This work was supported in part by the EU program
``Thales" (MIS 375734) and was also co-financed by the European
Union (European Social Fund, ESF) and Greek national funds under
the ``ARISTEIA II" Action. TNT would like to thank the Theory Unit
of CERN for its hospitality during the later stages of this work.
PS is grateful for financial support from the RFBR under grant
14-02-01092, as well as from the non-commercial ``Dynasty''
Foundation (Russian Federation).

\appendix

\section{Conventions}
\label{conventions}

Our convention for the Minkowski metric is $\eta_{\mu\nu} =
\eta^{\kn \mu\nu}={\rm diag}(+1,-1,-1,-1)$.

Fourier transforms are defined as:
\begin{align}\label{gr_pert9}
f(x)=\frac{1}{(2 \pi)^4}\int f(k) \, \e^{-i (kx)} d^{{\kern 1pt}4}
k\,, \qquad f(k)= \int f(x)\, \e^{i( k x)} d^{{\kern 1pt}4} x
\equiv \mathcal{F}[f(x)](k)\,.
\end{align}


We use the following symmetrization notation for tensorial
indices: $a_{(\mu}b_{\nu)}=[\vp
a_{\mu}b_{\nu}+a_{\nu}b_{\mu}]/2$\,.

\bigskip

{\bf Polarization tensors}. Given the radiation null wave-vector
$k^\mu$ and the null velocities $u^\mu$ and $u'^\mu$, define the
polarization vectors $e_1$ and $e_2$ by:
\begin{equation}
\label{vector_e1} e_1^\mu = \frac{1}{ k_{\bot} } \left[\frac{(ku)
{\kern 1pt}u'^\mu -(ku') {\kern 1pt}u^\mu}{2} -
\frac{ku-ku'}{ku+ku'}\,k^\mu \right], \qquad e_2^\mu =\frac{1}{2
k_{\bot} }\,\epsilon^{\kn \mu  \nu \lambda
\rho}u_{\nu}u'_{\lambda}k_{\rho}\,,  \end{equation} where
$k_{\bot}^2=(ku)(ku')$ and $\epsilon^{0123}=1$. They satisfy
\begin{equation} e_1\cdot e_2=k\cdot e_1=k\cdot e_2=0 \qquad {\rm
and} \qquad e_2\cdot u = e_2\cdot u'=e_2\cdot k=0\,.
\end{equation} In the center of mass frame of the collision under
study we have chosen $u^\mu=(1,0,0,1)$, $u'^\mu=(1,0,0,-1)$ and
$b^{\kn \mu}=(0,b,0,0)$. In addition, we have defined
$k^\mu=\omega(1, {\bf n}) \equiv \omega(1, \sin\vartheta
\cos\varphi, \sin\vartheta \sin\varphi, \cos\vartheta)$. Using
these, we obtain
\begin{equation} e_1^\mu=(0, \partial {\bf n}/\partial\vartheta)\,, \qquad
e_2^\mu=\frac{1}{\sin\vartheta}\,(0, \partial {\bf
n}/\partial\varphi) \end{equation} and \begin{equation} e_1 \cdot
u= \sin \vartheta \,, \qquad e_1 \cdot u'=-\sin \vartheta\,,
\qquad e_1\cdot b= -b{\kern 1pt}\cos \vartheta\, \cos\varphi
\,,\qquad e_2\cdot b=-b\sin\varphi\,. \end{equation} Finally,
using $e_1$ and $e_2$, we define the polarization tensors as
\begin{equation}
\ep_{+}^{\mu\nu}=\frac{e_1^{\mu}e_1^{{\vphantom{\mu}}\nu}-e_2^{\mu}e_2^{{\vphantom{\mu}}\nu}}{\sqrt{2}}\,
,\qquad  \ep_{\times}^{\mu\nu}=\sqrt{2}\, e_1^{(\mu}e_2^{\nu)}\,,
\qquad \ep_{\cp}^{\mu\nu} \eta_{\mu\nu}=0\,,  \qquad \cp = +,
\times\,.
\end{equation}

\bigskip

{\bf Useful integrals.} The following integrals are used in the main text:
\begin{align}\label{integrals}
&\int\limits_0^{2\pi} \sin^2\!\varphi \sin^2
\!\left(\frac{a\cos\varphi}{2}\right)\,d \varphi=\pi
\left(\frac{1}{2} -\frac{J_1(a)}{a}  \right), &\qquad&
  \int\limits_0^{2\pi} \cos^2\!\varphi \sin^2 \!\left(\frac{a\cos\varphi}{2}\right)\,d \varphi=\pi \left(\frac{1}{2} -J_0(a)+\frac{J_1(a)}{a}
  \right),\nn \\
   & \int\limits_0^{2\pi} \sin^4\!\varphi \sin^2 \!\left(\frac{a\cos\varphi}{2}\right)\,d
\varphi= 3\pi  \left(\frac{1}{8}   - \frac{ J_2(
  a)}{a^2}
 \right)\, ,&  &
  \int\limits_0^{2\pi} \cos^4\!\varphi \sin^2 \!\left(\frac{a\cos\varphi}{2}\right)\,d \varphi= \pi  \left[\frac{3}{8} +  \left(1 - \frac{3}{a^2}\right)
  J_2(
  a)\right].
 \end{align}

\section{Computation of  momentum integrals}
\label{Local_ints}

It is convenient for the computation of the integrals below to
define the four vector $b^{\kn \mu } $ by:
$$b^{\kn\mu}  \equiv\frac{1}{2}\left[\vph (bu)\, u'^\mu +(bu')\, u^\mu \right]+b_{\perp}^{\kn \mu} .$$

\textbf{Local integrals.} The basic scalar integral is
\begin{align}
I= \frac{1}{(2\pi)^{2}} \int \frac{ \delta(qu')\,
\delta(ku-qu)\,\e^{-i(qb)}}{q^2}\;\dq =
-\frac{\e^{-i(bu')(ku)/2}}{2{\kern 1pt}(2\pi)^{2}} \int
\frac{\e^{i\mathbf{qb}}}{\mathbf{q}^2}\,d^{\kn 2} \mathbf{q}=-
{\e^{-i(bu')(ku)/2}} \frac{\Phi(b)}{2}\
\end{align}
with $b\equiv \sqrt{-b\cdot b}$ and $\Phi(r)
\equiv-(1/2\pi) \ln (r/r_0)$.

The local vector momentum integral is defined as
\begin{align}
 I_{\mu}\equiv
\frac{1}{(2\pi)^{2}} \int \frac{ \delta(qu')\, \delta(ku-qu)\,
\e^{- i(qb)}}{q^2} \;q_{\mu} \; \dq = i\frac{\partial I}{\partial
b^{\kn \mu} } =-\e^{-i(bu')(ku)/2}
\left[\frac{(ku)\,\Phi(b)}{4}\,u'_\mu -i\frac{\Phi'(b)}{2b}\,
b_\mu \right].
\end{align}

The corresponding integrals for the primed particle are obtained
by the substitution $u^\mu \leftrightarrow u'^\mu, b^{\kn\mu} \to
-b^{\kn \mu}$.

\bigskip

\textbf{Stress integrals.}  The covariant stress integrals are
defined by
\begin{align}
J\equiv \frac{1}{(2\pi)^{2}}  \int \frac{ \delta(qu')\,
\delta(ku-qu)\,\e^{-i(qb)}}{q^2 (k-q)^2}\;\dq \,, \qquad
J_{\mu}\equiv i \frac{\partial J}{\partial b^{\kn\mu}} \,, \qquad
J_{\mu\nu} \equiv -\frac{\partial^2 J}{\partial b^{\kn \mu}
\partial b^\nu}\,.
\end{align}

\bigskip

$\bullet$ We start with the computation of $J$. We will then
compute the other stress integrals as derivatives of $J$,
following the same procedure as above, but to avoid unnecessary
complications we will give the final expressions only for the
special choice $b^{\kn \mu}=(0,1,0,0)$ made in the main text. We
obtain\kn\footnote{Use was made of the formulae
$$
 \int\limits_{0}^{\pi} e^{\pm i z \cos \theta}\sin^{2k}\theta d\theta =
 \frac{2^{k}\Gamma(k+1/2)\sqrt{\pi}}{z^k}J_k(z)\,, \qquad
 \int\limits_{0}^{\infty} dy \frac{y^{n+1}J_n(b y)}{[y^2+a^2]^{m}}=\frac{b^{\kn m-1} |a|^{n-m+1}}{2^{m-1}\Gamma(m)}\, K_{n-m+1}
 (b\hsp
 |a|)\,.
$$}
\begin{align}
J &= \frac{1}{8\pi^{2}} \int \frac{ \e^{ i \mathbf{q}\mathbf{b}
}}{\mathbf{q}^2 (\mathbf{k}-\mathbf{q})^2}\;d^{\kn 2}  \mathbf{q}
=\frac{1}{ 2\pi^{2}}\int\limits_0^1 dx\; \e^{-i (k b) x}
 \int\limits_0^{\pi} d \theta \int\limits_0^{ \infty} d q
\frac{ \e^{ i q b \cos \theta}}{[ {\mathbf{q}}^2 + \mathbf{k} ^2
x(1-x)]^2}\,  q   =\nn \\
&=\frac{1 }{ 4\pi } \int\limits_0^1 dx\; \e^{-i (k b) x}
\int\limits_0^{
 \infty}d {q}\,
\frac{ q \,J_{0}(qb)}{[ {\mathbf{q}}^2 + \mathbf{k} ^2 x(1-x)]^2}
  =
 \frac{b^{\kn 2}}{8\pi} \int\limits_0^1 dx \,\e^{-i (k
b) x} \hat{K}_{-1} \!\left(k_\perp b \sqrt{x(1-x)} \right), \nn
\end{align}
where $k_\perp \equiv |{\bf k}| = \sqrt{(ku)(ku')}$ is the
magnitude of the two-dimensional transverse part of $k^{\kn\mu}$.

Also\kn\footnote{A useful formula is \;$ \ds J^{0}= - J^{z}
=\frac{(ku)}{2}\,J\,. $},
\begin{align}
\label{Jm}
 J_\mu =\frac{b^{\kn 2}}{8\pi}\int\limits_0^1
dx \;\e^{-i (k  b) x} \left[N_\mu    \hat{K}_{-1} \!\left(\zeta
\right)+ i \frac{b_\mu  }{b^{\kn 2}}\, {K}_{0}\! \left(\zeta
\right) \right]\,,\;\;\; \;\;\; N^\mu \equiv x k^{\kn \mu}
-\frac{1}{2}\left[x(ku')\,u^\mu -(1-x) (ku)\,{u'}^\mu \vph\right],
\end{align}
where $\zeta  \equiv k_{\perp} b \sqrt{x(1-x)}$\,.

\bigskip
$\bullet$ Next consider $(k  J)$. With the on-shell condition
$k^2=0\,$ it can be represented as
\begin{align}
\label{chr1} k\cdot J & =   \int \frac{ \delta(qu')\,
\delta(ku-qu)\, \e^{- i(qb)}}{q^2 (k-q)^2} \;(k  q) \; \dq
=\frac{1}{2}  \int \frac{ \delta(qu')\, \delta(ku-qu)\, \e^{-
i(qb)}}{q^2 (k-q)^2} \,
\left[q^2 -(k-q)^2\right]\, \dq =\nn \\
& =\frac{1}{2} \left[
\e^{- i(kb)}I^* -I \right]=  \frac{I}{2} \left[ \e^{- i(kb)}- 1
\right]=\frac{ \Phi(b)}{4} \left[1- \e^{- i(kb)} \right].
\end{align}

On the other hand, the integral form
\eqref{Jm} of $J_\mu$ becomes
\begin{align}
 k\cdot J =\frac{1}{8\pi}\int\limits_0^1 dx\; \e^{-i
(k b)x} \left[ \left(\frac{1}{2}-x  \right){k}_{\perp}^{2}b^{\kn
2} \hat{K}_{-1}\! \left(k_\perp b \sqrt{x(1-x)} \right)+ i (kb)
\,{K}_{0} \!\left(k_\perp b \sqrt{x(1-x)} \right) \right],\nn
\end{align}
whose integrand is the derivative of $-\e^{-i (k b)x} {K}_{0}
\!\left(k_\perp b \sqrt{x(1-x)} \right) $. Hence,
\begin{align}
\label{kJ}
 k\cdot J  =\frac{1}{ 8\pi } \left.\e^{-i (k b)x}
\hat{K}_{0} \!\left(k_\perp b \sqrt{x(1-x)} \right)
\right|_1^0=\lim_{d\to+0} \frac{\hat{K}_{ d/2}(0)}{8\pi} \left[1-
\e^{- i(kb)} \right] =\frac{ \Phi(b)}{4 } \left[1- \e^{- i(kb)}
\right].
\end{align}

\bigskip

$\bullet$ Following similar steps, we find for the tensorial integral
\begin{align}\label{iij1}
 J_{\mu\nu} = \frac{1}{8\pi} \int\limits_0^1
dx\;\e^{-i(kb)x} & \left[ b^{\kn 2} N_{\mu} N_{\nu}
\hat{K}_{-1}(\zeta)
  + \left(2iN_{(\mu }
b_{\nu )} -\eta_{\mu\nu}+ u^{\vphantom{\prime}}_{(\mu }u'_{\nu )}
\right) {K}_{0}( \zeta ) - \frac{b_{\mu}b_{\nu}}{b^{\kn 2}
}\,\hat{K}_{1}( \zeta )\right].
\end{align}

\bigskip

$\bullet$ Its trace $\Sp J\equiv \eta^{\kn \mu\nu} J_{\mu\nu}$ is
given by
\begin{align}
\Sp J  = \frac{1}{8\pi} \int\limits_0^1 dx\;\e^{-i(kb)x} &
\left[-x^2 b^{\kn 2} k_{\perp}^2\hat{K}_{-1}(\zeta)+ 2i x (k b)
{K}_{0}( \zeta )+ \hat{K}_{1}( \zeta )- 2{K}_{0}( \zeta )\right].
\nn
\end{align}
Make use of the Macdonald-function identity $
\hat{K}_{1}(z)=z^2\hat{K}_{-1}(z)$ to obtain
\begin{align}
\Sp J  =\frac{1}{8\pi} \int\limits_0^1 dx\;\e^{-i(kb)x} &
\left[-x(2x-1)\, b^{\kn 2} k_{\perp}^2\hat{K}_{-1}(\zeta)+
2\left(\vph i x (k b) -1\right) \!{K}_{0}( \zeta )\right]. \nn
\end{align}
The integrand is the total derivative of $-2x\,\e^{-i (k b)x}
{K}_{0} \!\left(k_\perp b \sqrt{x(1-x)} \right)$. Hence
\begin{align}
\label{SpJ}
\Sp J  =\frac{x}{ 4\pi} \left.\e^{-i (k b)x} {K}_{0}
\!\left(k_\perp b \sqrt{x(1-x)} \right) \right|_1^0=-\e^{-i (k b)}
\frac{\Phi(b)}{2}=\e^{-i (k b)} I\,.
\end{align}

\section{Asymptotic expansion  of higher-frequency integrals}
\label{Lm}

We are interested here in the expansion of the typical integrals
$$L_{m}\equiv \int\limits_0^1 \e^{-icx} \hat{K}_m \!\left(a \sqrt{x(1-x)}\right)dx$$
(where $c= k\cdot b =-a\cos \varphi$) in powers of $1/a$ for
\mbox{$a\gg 1$}. The presence of the $\sqrt{x(1-x)}$ in the
argument of the $K_m$, prevents the direct use of an asymptotic
expansion of the Macdonald function itself. Instead, we shall
proceed as follows: we represent
$$ L_{m} = \sqrt{\frac{2}{\pi}} \,\left(\frac{a}{2}\right)^{2m+1}
\int\limits_0^{\infty} dy  \check{K}_{m+1/2} \!\left(\frac{a}{2}
\sqrt{y^2+1} \right)\int\limits_0^{1}\e^{-icx}\ch\!\left(\vph ay
 \left(x- {1}/{2} \right)\right)dx\,, $$
where $\check{K}_m(z)\equiv z^{-m} K_m(z)$. Making use of
$$ \int\limits_0^{1}\e^{-icx}\ch\!\left(\vph s
 \left(x- {1}/{2} \right)\right)dx  = \frac{2 {\kern 1pt}\e^{-ic/2}}{s^2 +c^2} \left[
s \sh \frac{s}{2} \cos \frac{c}{2}+ c\ch \frac{s}{2} \sin
\frac{|c|}{2}
 \right], \qquad s^2 \geqslant c^2,$$
one obtains
 \begin{align}\label{naxos}
 L_{m} =\frac{2 {\kern 1pt}\e^{-ic/2}}{a^2}
 \sqrt{\frac{2}{\pi}}
\,\left(\frac{a}{2}\right)^{2m+1} \int\limits_0^{\infty} dy\,
\frac{\check{K}_{m+1/2} \!\left(\ds\frac{a}{2} \sqrt{y^2+1}
\right)}{ y^2 +\cos^2\! \varphi} \left[ ay \sh \frac{ay}{2} \cos
\frac{c}{2}+c \ch \frac{ay}{2} \sin \frac{|c|}{2}
 \right].
 \end{align}

Then one may use the expansion
$$\frac{1}{y^2+\cos^2\! \varphi} = \sum_{k=0}^{\infty} \frac{ \sin^{2k}\!\varphi}{ (y^2+1)^{k+1}} \,,$$
which converges since $y^2+1 \geqslant 1$, $
\sin^{2}\!\varphi\leqslant 1 $\kn\footnote{Notice, that the series
is convergent even for the limiting value $ \cos \varphi=0$, while
the integral \eqref{naxos} converges for $y\to 0$, as a
consequence of  $$\lim_{y\to 0}\frac{ \sh(ay/2)}{y}=\frac{a}{2}\,,
\qquad  \lim_{c\to 0} \frac{c \sin (|c|/2)}{y^2+c^2/a^2} =
\frac{a^2}{2}\,.$$}.

The next step is to integrate over $y$ using the formulae
($q\geqslant |r|$)
\begin{align}
\label{lead0} q^{2\nu}\int\limits_0^\infty \! dy \,
\check{K}_{\nu} \! \left(q\sqrt{y^2+1}\right)  \left\{
\arraycolsep=0cm\begin{array}{c}
y \sh(ry)  \\
\ch(r y) \\
\end{array}
 \right\}= \sqrt{\frac{\pi}{2}} \left\{\arraycolsep=0cm
\begin{array}{c}
r \hat{K}_{\nu-3/2}  (\sqrt{q^2-r^2}  ) \\
\hat{K}_{\nu-1/2} (\sqrt{q^2-r^2} )  \\
\end{array}
 \right\}
\end{align}
together with the identity
\begin{align}
\label{Mac_reduce}
K_{\nu}(z)=K_{\nu+2}(z)-\frac{2(\nu+1)}{z}\,K_{\nu+1}(z) \simeq
K_{\nu+2}(z) \,, \quad{\rm for} \; \;z\gg 1\,.
\end{align}

Explicitly, applying (\ref{Mac_reduce}) once, the leading term is the one containing $\sh(ay/2)$ and is given by:
$$  \frac{2 {\kern 1pt} }{a}  \sqrt{\frac{2}{\pi}}
\,\left(\frac{a}{2}\right)^{2m+3} \int\limits_0^{\infty} dy\,
 \check{K}_{m+5/2} \!\left(\frac{a}{2} \sqrt{y^2+1} \right)  y \sh \frac{ay}{2} \cos \frac{c}{2} = \frac{2^{m+2}\Gamma(m+1)}{a^2}\,\cos\frac{c}{2} \,. $$

The first subleading term  is $\mathcal{O}\kn(a^{-3})$: it comes
from $\ch (ay/2)$ times basic Macdonald; the next subleading terms
are $\mathcal{O}\kn(a^{-4})$: they come (i) from $y\sh(ay/2)$
times previous shift-index term; and (ii) $y\sh (ay/2)$ times
first correction in $\sin^2\!\varphi$. The end result is:
$$ L_m = \frac{2^{m+2}\Gamma(m+1)}{a^2} \,\e^{-ic/2} \cos\frac{c}{2} \left[1- \frac{m+1}{a^2} \left( 4\kn  a\cos\varphi \, \tg\frac{c}{2}
+8 {\kern 1pt}(2m+3)
 - 16  (m+2) \,\sin^2\!\varphi \right)+ \mathcal{O}\Bigl(\frac{1}{a^3}\Bigr)   \right] . $$

\end{document}